\documentclass[a4paper,11pt]{article}
\pdfoutput=1 

\usepackage{jheppub} 
\usepackage{textcomp,parskip,amsmath,physics,caption,subcaption,color}

\makeatletter
\def\@fpheader{\relax}
\makeatother
\renewcommand\footnoterule{\kern-2pt \hrule width 6in \kern 2.6pt}
\title{Holographic Description of Noncommutative Schwinger Effect}

\author[a]{Udit Narayan Chowdhury. }

\affiliation[a]{Saha Institute of Nuclear Physics, Block-AF, Sector-1, Salt Lake.\\
 Kolkata 700064, India

and

Homi Bhabha National Institute, Training School Complex, 
Anushakti Nagar.\\
Mumbai 400085, India}

\emailAdd{udit.chowdhury@saha.ac.in, chowdhury.udit@gmail.com}

\abstract{We consider the phenomenon of spontaneous pair production in presence of an external electric field for noncommutative Yang Mills theories.
Using Maldacena's holographic conjecture the threshold electric field for pair production is computed from the quark/anti-quark potential for noncommutative theories. 
As an effect of noncommutativity, the threshold electric field is seen to be smaller than its commutative counterpart. We also estimate the correction to the
production rate of quark/anti-quark pairs to first order of the noncommutative deformation parameter. Our result bears resemblance with an earlier related work 
(based on field theoretic methods).}

\begin{document} 
\maketitle
\flushbottom

\section{Introduction}
\label{sec:intro}

Quantum Field Theory is primarily studied in its perturbative regime. However there exists quite some novel non-perturbative features of quantum field
theories amongst which the Schwinger Effect \cite{Schwinger:1951nm} stands its ground\footnote{For a recent review see \cite{Gelis:2015kya}}. The vacuum of Quantum Electrodynamics is a bath of $e^{+} e^{-}$ virtual
pairs which gets created and annihilated instantaneously. However
in presence of an external electric field, the $e^{+}$ $e^{-}$ pairs spontaneously become real and their production rate in the weak-coupling weak field approximation
is given by \cite{Schwinger:1951nm}.
\begin{equation}
\label{sdecay}
\Gamma=\frac{(eE)^{3}}{(2\pi)^{3}}e^{\frac{-\pi m^{2}}{eE}}
\end{equation}
This expression holds for weak coupling and weak electric fields only. The exponential suppression hints that the pair production process can be modeled as quantum mechanical tunneling in a
certain potential barrier.
For the electron-positron pairs  to become real they should gain at-least an energy equal to sum of 
their rest masses ($2m$). However in presence of an external electric field the virtual particles  gather an energy of $eEx$  via electromagnetic interactions, $x$ being their separation distance. To
understand the situation better let us assume that the positron is located distance $-\frac{1}{2}x$
and the electron at distance $\frac{1}{2}x$ from the origin along the direction parallel to the electric field. To become physical particles,  the electron has to climb through a potential barrier 
of height $m$ (same for the positron) by gaining energy from the external electric field.
If the virtual pairs are separated by a distance $x_{*}$ such that $\frac{1}{2}eEx_{*}=m$ then the electron (and the positron) become real particle as it now has the required rest mass energy. This
value $x_{*}=\frac{2m}{eE}$ is the width of the potential barrier. Thus the transmission 
co-efficient is approximately $\text{exp}(-x_{*}\sqrt{2m.m})\sim  e^{-\frac{4m^{2}}{eE}}$. This is exactly the content of the Schwinger formula \eqref{sdecay}.

Motivated by the this analogy, let us look from the perspective of a "virtual" $q\bar{q}$ dipole. When the "virtual" $q\bar{q}$ dipole has a separation $x$, the total effective potential barrier they
encounter can be estimated to be  of the form 
\begin{align}
V_{\text{total}}(x)=2m-\frac{\alpha}{x}-eEx
\end{align}
In this picture the virtual particles become real by tunneling through the above said potential barrier. The first two terms indicate the self energy including the coulombic interaction between  $q\bar{q}$ pairs.
For small separation the coulomb term dominates the expression and the potential is negative. At large values of $x$ the effect of electric field takes dominance making the potential negative too. 
For small electric field i.e. $E< \frac{m^{2}}{e\alpha}$ there exists two zero points of the potential profile and the potential is positive in intermediate regimes of separation $x$. In this case the 
particles become real by tunneling through
this barrier and the production rate is exponentially suppressed as described by the Schwinger formula. However for electric fields $E> \frac{m^{2}}{e\alpha}$, the potential becomes negative all along 
and ceases to put up a barrier,
indicating a catastrophic instability of  vacuum where the $q\bar{q}$ are spontaneously produced.
The value of electric field for which the potential changes its character is called the "threshold electric field" $\mathcal{E}_{T}$.  

The idea of noncommutative quantum field theories \cite{PhysRev.71.38}\cite{Szabo:2001kg}, where spacetime position, viewed as operators itself ceases to commute was originally 
proposed to curb the UV divergences appearing in interacting quantum field theories. The idea received revival when some noncommutative field theories were
found to be effective low energy limit of open string theories on a Dp brane with a constant NS-NS two-form $B_{\mu\nu}$, the noncommutative feature being a dynamical 
consequence of quantization \cite{Schomerus:1999ug}\cite{Seiberg:1999vs}. However
the low energy limit of these string
theories turn out to be quantum field theories defined on a "noncommutative" spacetime. There exists noncommutative generalizations of Riemannian geometry \cite{Madore:2000aq}
on which the standard model can be defined, wherein the parameters of the theory are interpreted as geometric invariants. Mathematically this amounts to abandoning the smooth structure of spacetime  in favor
of a space equipped with 
a noncommutative algebra of real valued functions much like the transition from classical to quantum physics via phase-space methods. 
The transition from commutative theories to its noncommutative 
counterpart along the lines stated above is achieved via replacing the ordinary product between functions by the Moyal/star product($\star$).
\begin{align}
 F(x)G(x)\longrightarrow F(x) \star G(x) = \exp \big(\;\frac{i}{2}\theta^{\mu\nu}\partial_{\mu}\partial^{\prime}_{\nu}\;\big)F(x)G(x^{\prime})\Big\vert_{x=x^{\prime}}
\end{align}
The above equation implies $[x^{\mu},x^{\nu}]_{\star}=x^{\mu} \star x^{\nu} - x^{\nu} \star x^{\mu}=i\theta^{\mu\nu}$, signifying non-vanishing commutation relations between spacetime 
coordinates (viewed as operators) itself.  Noncommutative quantum field theories are generically Lorentz violating (due to presence of a noncommutative 
parameter $\theta_{\mu\nu}$), the effects of which
 are small to be detected in practical experiments with current experimental bound on the noncommutative parameter to be around  $(\mid \theta_{\mu\nu} \mid \lesssim (10 \text{TeV})^{-2} )$ by conservative estimates
 \cite{Carroll:2001ws}. 

The Schwinger Effect in noncommutative QED has been calculated in \cite{Chair:2000vb} where 
a correction to the pair production rate has been found leading to a decrement in the threshold electric field as a consequence of noncommutativity. However to 
carry on the same kind of analysis for strong coupling in general becomes an uphill task and the presence of noncommutativity makes matters worse. 
The Gauge/Gravity (holographic) correspondence \cite{Aharony:1999ti} which links a strongly coupled gauge
theory to classical gravity is an important tool in these kind of scenario. The Schwinger mechanism has been argued in the holographic setting in \cite{Semenoff:2011ng}. It has also been shown via 
holographic methods \cite{Sato:2013dwa} that 
Schwinger effect in a large N confining gauge theory admits a "new kind" of critical electric field (apart from the usual threshold value). If the external electric field is less than the confining
string tension $\sigma_{str}$, the pair production is prohibited as the 
effective potential barrier remains positive instead of dumping out at large $q\bar{q}$ separation. However when the electric field is larger than $\sigma_{str}$, pair production is allowed as a tunneling
process. Thus, at this value ($\sigma_{str}$) of electric field
a confinement/deconfinement transition happens. 

In this paper we like to study the Holographic Schwinger effect for "quarks" coupled to a large N noncommutative gauge theory in presence 
of external U(1) gauge field. The large N noncommutative gauge theory (NCYM) is realized by it's relevant holographic dual geometry (to be mentioned later) while the quark are modeled as massive strings extending
from the interior to a large but finite position in the holographic direction so that the  mass of the quarks is not infinite. We evaluate the inter-quark potential (both numerically and
analytically when possible) for the large N NCYM from the rectangular Wilson loop by calculating the extremal area of a string worldsheet ending along a rectangular contour at the boundary. The effective potential 
for describing Schwinger effect by is then calculated by introducing an external electric field. We analytically find out the value of the external electric field for which the effective potential puts up a tunneling
behavior i.e. the threshold electric field. We repeat the same calculation for finite temperature NCYM and observe that the thermal contribution to the threshold electric field don't mix up up the noncommutative ones.
We proceed to find out the decay rate by finding out the on-shell value of the Polyakov action coupled to an external electric field for a string with circular contour at the boundary. We use perturbations over the known
result of circular holographic Wilson loop of ordinary YM and hence find out the correction of the Schwinger decay rate \cite{Semenoff:2011ng} upto first order of the noncommutative deformation parameter. 

This paper is organized as follows. In section 2 we review the derivation of Schwinger effect in a super-conformal SU(N) gauge theory by relating the same to the 
expectation value of circular Wilson loop. Also in the same section the basics of noncommutativity in string theory is reviewed for the sake of clarity. Section 3 
is devoted to the analysis of effective potential for virtual particles in NCYM plasma, from where the explicit form of the threshold electric field is found out by
analytical means. In section 4 we compute the first order noncommutative correction to 
the pair production rate of Schwinger particles. We close this paper by conclusions in section 5.
\section{Pair Production and Noncommutativity}
\subsection{Pair Production in SYM and the Wilson Loop}
\label{pair production}
In its most prominent avatar \cite{Aharony:1999ti}\cite{Maldacena:1997re}, the holographic duality relates $\mathcal{N}=4$ Super Yang Mills theory to type IIB String 
Theory in $AdS_{5}$ \texttimes \,$S^{5}$. To study Schwinger effect one has to account for "massive" matter (corresponding to a probe brane) in fundamental
representation and an U(1) gauge field. The way to do so is to break the symmetry group of the problem from SU(N+1) to SU(N) \texttimes \:U(1) with the Higgs mechanism.
Such methods were first introduced in \cite{Drukker:1999zq}, the following closely follows \cite{Affleck:1981bma}\cite{Sato:2013pxa}. For more sophisticated treatment
refer to \cite{Gordon:2014aba}\cite{Gordon:2017dli}. The bosonic part of $\mathcal{N}=4$ SYM for the SU(N+1) theory in euclidean signature reads as.
\begin{align}
\label{sym} 
\hat{S}^{SU(N+1)}=\frac{1}{g_{YM}^{2}} \int d^{4}x\;\Big(\;\frac{1}{4} \hat{F}_{\mu\nu}^{2} + \frac{1}{2}(\hat{D}_{\mu}\hat{\Phi}_{i})^{2} 
-\frac{1}{4}[\hat{\Phi}_{i},\hat{\Phi}_{j}]^{2}\;\Big) 
\end{align}
$\hat{F}_{\mu\nu}$ is the field strength of the SU(N+1) gauge field $\hat{A}_{\mu}$. $\hat{\Phi}_{i}\;(i=1,...,6)$ collectively denotes six scalars in the adjoint 
representation of SU(N+1). The gauge group is broken as, 
\begin{align}
 \label{decomp}
 \hat{A}_{\mu}\rightarrow
\begin{pmatrix}
 A_{\mu} & \omega_{\mu} \\
 \omega_{\mu}^\dag & a_{\mu}  
 \end{pmatrix}\;\;\;\;,\;\;\;\;
 \hat{\Phi}_{i}\rightarrow
 \begin{pmatrix}
\Phi_{i} & \omega_{i}\\
\omega_{i}^\dag & m\phi_{i}
 \end{pmatrix}
\end{align}
The non-diagonal parts, $\omega_{\mu}$ and $\omega_{i}$ 
transform in the fundamental representation of SU(N) and form the so called W-boson multiplet. The VEV of the SU(N+1) scalar fields is supposed to be of the form,
\begin{align}
\label{vev}
 \expval{\hat{\Phi}_{i}}=\text{diag}(0,...,0,m\phi_{i})\;\;\;\;;\;\;\;\;\sum_{i=1}^{6} \phi_{i}^{2}=1
\end{align}
As a result of the decomposition \eqref{decomp} the SU(N+1) action \eqref{sym} breaks up into three parts of the following form \cite{Drukker:1999zq}
\begin{align}
 \label{brkup}
 \hat{S}^{SU(N+1)} \longrightarrow\;\;\; S^{SU(N)} + S^{U(1)} + S_{W}
\end{align}
 $S^{U(1)}$ is basically the free QED action constructed out of the gauge field $a_{\mu}$.
 $S_{W}$ governs the dynamics of the W bosons and its coupling to the gauge fields. Disregarding the $\omega_{\mu}$'s \footnote{The $\omega_{\mu}$'s start coupling to
 the $\omega_{i}$ via $\omega_{\mu}D_{\mu}\omega_{i}$. The effect of 
 these couplings to the vacuum energy density of the $\omega_{i}$ are damped by $\frac{1}{N}$ at least. Thus in the large N limit, 
 the $\omega_{\mu}$ are neglected for the present study.}  and higher order terms the W boson action reads,
\begin{align}
 \label{sw}
 S_{W}= \frac{1}{g_{YM}^{2}} \int d^{4}x \;\Big[\;\arrowvert D_{\mu} \omega_{i}\arrowvert^{2} + \omega_{i}^\dag (\Phi_{j}-m\phi_{j})^{2} \omega_{i} -
 m^{2}\omega_{i}^\dag\phi_{i}\phi_{j}\omega_{j} +...\;\Big]
\end{align}
$D_{\mu}$ is equipped both with the SU(N) gauge field $A_{\mu}$ and also with the U(1) gauge field $a_{\mu}$ i.e. $D_{\mu}=\partial_{\mu}-iA_{\mu}-ia_{\mu}$.
By expanding the action $S_{W}$ and choosing $\phi_{i}=(0,0,0,0,0,1)$, the mass term for $\omega_{6}$ vanishes while those for $\omega_{i}$, $i=1,...,5$ remain. 
For the present scenario the SU(N) gauge field $A_{\mu}$ 
is a dynamical field and the U(1) gauge field $a_{\mu}$ is a "fixed external" field of the form $a_{\mu}=a^{(E)}_{\mu}=-Ex_{0}\delta_{\mu1}$. The external gauge field contributes to the vacuum energy density
via the covariant
derivative in $S_{W}$ as mentioned above. The pair production rate is given by the imaginary part of vacuum energy density \cite{Itzykson:1980rh}. 
\begin{align}
\label{dform}
\Gamma&= -2\;\text {im ln} \int \mathcal{D}A \mathcal{D}\Phi \mathcal{D}\omega\;\; e^{-S^{SU(N)}-S_{W}} \nonumber \\
&\approx \;5\text{N}\;\text {im} \int \mathcal{D}A \mathcal{D}\Phi \;e^{-S^{SU(N)}}\text{tr}_{SU(N)}\;\hat{\text{tr}}\;\text{ln} (-D_{\mu}^{2}+(\Phi_{i}-
m\phi_{i})^{2})
\end{align}
The factor of N comes form the number of $\omega_{i}$ s. By using Schwinger's parametrization and world-line techniques \cite{Schubert:2001he} , one can express the pair production rate  \eqref{dform} as a path integral for
a particle subject to an appropriate Hamiltonian under boundary conditions, ${x(\tau=0)=x(\tau=T)}$
\begin{align}
 \label{wqft}
 \Gamma=-5\text{N}\;\text{im}\expval{\text{tr}_{SU(N)}\mathcal{P} \int_{0}^{\infty} \frac{dT}{T} \int \mathcal{D}x(\tau)\;
 e^{-\int_{0}^{T}d\tau\Big[\frac{1}{4}\dot{x}^{2}+iA_{\mu}\dot{x}_{\mu}+ia^{(E)}_{\mu}\dot{x}_{\mu}+(\Phi_{j}-m\phi_{j})^{2} \Big] }}
\end{align}
Using saddle point approximations as in \cite{Affleck:1981bma}\cite{Sato:2013pxa} and assuming the mass to be heavy i.e. $m^{2}>>E$ , \eqref{wqft} becomes 
proportional to the path integral of the particle subjected to the "external" gauge field $a^{(E)}_{\mu}$ times a phase factor, namely the SU(N) Wilson loop.
\begin{align}
 \label{rateloop}
 \Gamma \sim& -5\text{N}\;\int \mathcal{D}x \;\text{exp}\Big(-m\int_{0}^{1}d \tau\;\sqrt{\dot{x}^{2}}+
 i\int_{0}^{1}d \tau \;a^{(E)}_{\mu}\dot{x}_{\mu}\Big)\expval{W[x]}
 \end{align}
 \begin{align}
 \label{wloopqft}
 &\expval{W[x]}=\expval{\text{tr}_{SU(N)}\mathcal{P}\text{exp}\Big( \int_{0}^{1}d \tau \big(iA_{\mu}\dot{x}_{\mu}+
 \Phi_{j}\phi_{j}\sqrt{\dot{x}^{2}}\;\big) \Big)}_{SU(N)}
\end{align}
In the above $\mathcal{P}$ indicates the path ordering of the time parameter $\tau$ and the SU(N) nonabelian index is handled with  matrix trace, $\text{tr}_{SU(N)}$. 
Evaluating \eqref{rateloop} by the method of steepest descent the "classical" trajectory becomes a circle . 
So the production rate is proportional to the expectation value of the circular Wilson Loop and can be computed via holographic
conjecture in the large N limit.

From the expression \eqref{rateloop}, it naively seems that the pair production rate is nonzero in general implying the vacuum energy (of an SU(N) gauge theory) density to have an imaginary part 
 even in absence of an external electric field . However this is not the case. The expression in front of the  Wilson loop in \eqref{rateloop} can be evaluated by steepest descent methods and leads to a multiplicative
 factor of $\frac{E^{2}}{(2\pi)^{3}}$. Thus the pair production rate ( and the imaginary part of vacuum energy density) indeed goes to zero when the external electric field vanishes. A detailed calculation 
 of the same is presented in \cite{Affleck:1981bma}\cite{Gordon:2014aba}.
\subsection{Noncommutativity from String Theory}
The effective worldsheet action in presence of $B_{\mu\nu}$ field is given by,
\begin{align}
\label{stringaction}
 S=\frac{1}{4\pi \alpha^{\prime}}\int_{\Sigma} d^{2}s \Bigg[\partial_{a}X^{\mu}\partial^{a}X^{\nu}\eta_{\mu\nu}+\varepsilon^{ab}\partial_{a}X^{\mu}
 \partial_{b}X^{\nu}B_{\mu\nu}\Bigg]
\end{align}
The equations of motions along with the boundary conditions when $dB=0$ are
\begin{align}
 \label{ws}
 &\big(\partial_{t}^{2}-\partial_{s}^{2} \big)X^{\mu}(t,s)=0\\
 \label{bc}
 &\partial_{s}X^{\mu}(t,s)+B_{\nu}^{\;\mu}\cdot\partial_{t}X^{\nu}(t,s)\Big\vert_{s=0,\pi}=0
\end{align}
The boundary conditions \eqref{bc}  are neither Neumann nor Dirichlet, one can indeed try to diagonalize \eqref{bc} to be Neumann like by 
redefining the fundamental variables leading to the so called "open string metric" \cite{Seiberg:1999vs} . However a more direct attack
along the lines of \cite{Chu:1998qz} is to solve the equation of motion \eqref{ws} first and constrain the solution by \eqref{bc}. For
$B=B_{23}\;dX^{2}\wedge dX^{3}$, the solution of \eqref{ws} compatible with \eqref{bc} is
\begin{align}
\label{mode}
 X^{2}(t,s)=q^{2}_{(0)}+\big(a^{2}_{(0)}t+a^{3}_{(0)}B_{23}\;s\big)+\sum _{n\neq 0}\frac{e^{-int}}{n}\Big(ia^{2}_{(n)}\cos ns+a^{3}_{(n)}B_{23}\;\sin ns \Big) 
\end{align}
A similar solution accompanies $X^{3}(t,s)$,  the forms of which encode the  nontrivial boundary conditions. The solutions for the other co-ordinates are the 
usual ones \cite{Blumenhagen:2013fgp}. The canonical momentum of the action \eqref{ws} is by the virtue of mode expansion \eqref{mode},
\begin{align}
 \label{cannmomentum}
 \Pi^{2}(t,s)=&\frac{1}{2\pi \alpha^{\prime}}\Big(\partial_{t}X^{2}(t,s)-B_{23}\;\partial_{s}X^{3}(t,s)\Big)\nonumber\\
 =&\frac{1}{2\pi \alpha^{\prime}}\Big( a^{2}_{(0)}+\sum_{n\neq 0}a^{2}_{(n)}\;e^{-int}\cos ns\Big)\;\Big(1+(B_{23})^{2}\Big)
\end{align}
The fact that the current scenario leads to noncommutativity was first recognized in \cite{SheikhJabbari:1997yi}\cite{Ardalan:1998ce} as the canonical momentum 
at the ends of the string become functions of the spatial derivatives of the string coordinates as per \eqref{bc} and \eqref{cannmomentum}. From the symplectic $2$-form
the canonical commutation relation of the modes are,
\begin{align}
 \label{commutationrelations1}
&\big[q^{2}_{(0)},q^{3}_{(0)}\big]= i \frac{2\pi \alpha^{\prime}B_{23}}{1+(B_{23})^{2}}\\
 \label{commutationrelations2}
&\big[q^{2}_{(0)},a^{2}_{(0)}\big]=\big[q^{3}_{(0)},a^{3}_{(0)}\big]=i\frac{2\alpha^{\prime}}{1+(B_{23})^{2}}\\
 \label{commutationrelations3}
&\big[a^{2}_{(-n)},a^{2}_{(n)}\big]=\big[a^{3}_{(-n)},a^{3}_{(n)}\big]=\frac{2n\alpha^{\prime}}{1+(B_{23})^{2}}
 \end{align}
From the mode expansion \eqref{mode} and the relations \eqref{commutationrelations1}-\eqref{commutationrelations3} one has
\begin{align}
 \label{blabla}
\big[X^{2}(t,s),X^{3}(t,s^{\prime})\big]=i\frac{2\alpha^{\prime}B_{23}}{1+(B_{23})^{2}}\bigg[(\pi-s-s^{\prime})-\sum_{n\neq0}\frac{1}{n}\sin n(s+s^{\prime})\bigg]
\end{align}
The second term in \eqref{blabla} sums up to zero when $s+s^{\prime}=0,2\pi$. Therefore the end points of the string become noncommutative. The nontrivial part of the 
normal ordered Virasoro constraints and the total momentum which accompanies \eqref{stringaction} are given by
\begin{align}
 \label{virasoro}
 &L_{n}=\frac{1}{4\alpha^{\prime}}:\sum_{m}\Big[\big(1+(B_{23})^{2}\big)\big(a^{2}_{(n-m)}a^{2}_{(m)}+a^{3}_{(n-m)}a^{3}_{(m)}\big)+\sum_{i,j\neq 2,3}\eta_{ij}
 a^{i}_{(n-m)}a^{j}_{(m)}\Big]:\\
 \label{momentum}
 &P^{2}_{total}=\frac{1}{2\alpha^{\prime}}a^{2}_{(0)}\Big(1+(B_{23})^{2}\Big)\;\;\;\;;\;\;\;\;P^{3}_{total}=\frac{1}{2\alpha^{\prime}}a^{3}_{(0)}\Big(1+(B_{23})^{2}
 \Big)
\end{align}
It is clear from the above that the mass of the particle becomes dependent on the value of the $B_{\mu\nu}$ field. However the noncommutative field theories constructed 
out of Moyal product leave the mass of the particle (quadratic part of the Lagrangian) unchanged. Since the string equations of motion and boundary conditions
are linear equations, one can redefine the operators to be in terms of which the mass of the theory remains unaltered \cite{Bilal:2000bk}.
\begin{align}
 \label{redifined}
 \hat{q}^{2,3}_{(0)}=\sqrt{1+(B_{23})^{2}}\;{q}^{2,3}_{(0)}\;\;;\;\;\hat{a}^{2,3}_{(n)}=\sqrt{1+(B_{23})^{2}}\;{a}^{2,3}_{(n)}
\end{align}
It terms of the which the only nontrivial commutation relation become
\begin{align}
 \label{moyal}
 \big[\hat{q}^{2}_{(0)},\hat{q}^{3}_{(0)} \big]=2\pi i\alpha^{\prime}B_{23} \equiv i\theta
\end{align}
It has been checked that perturbative string theory in the present backdrop corresponds to Noncommutative Yang Mills at one loop. For further details see
\cite{Bilal:2000bk} and references there-within.
\section{Potential Analysis of Noncommutative Schwinger Effect}
We start with a brief description of the holographic dual of Noncommutative Yang Mills (NCYM)\cite{Maldacena:1999mh}\cite{Alishahiha:1999ci}. In the spirit of AdS/CFT 
correspondence one looks for supergravity solutions with a non zero asymptotic value of the B field. Such a solution is the D1-D3 solution which in the string frame
looks like,
\begin{align}
\label{d1-d3}
&\;\;\;ds^2_{str}=\frac{1}{\sqrt{F}}\;[-dx^{2}_{0}+dx^{2}_{1}+H(dx^{2}_{2}+dx^{2}_{3})\;]+\sqrt{F}[dr^{2}+r^{2}d\Omega^{2}_{5}]\nonumber\\
&\;\;\;\;\;\;\;\;\;\;\;\;\;\;\;F=1+{\alpha^{\prime}}^{2} \frac{R^{4}}{r^{4}}\;\;\;\;\;\;;\;\;\;\;\;\;H=\frac{\sin ^{2} \psi}{F}+\cos ^ {2}\psi\nonumber\\
&\;\;\;\;\;\;\;B=\frac{H \tan \psi}{F} dx_{2}\wedge dx_{3}\;\;\;\;;\;\;\;\;\;\;e^{2 \phi}=g_{s}^{2} H 
\end{align}
The solution \eqref{d1-d3} is asymptotically flat and represents $N_{(1)}$ D1 branes dissolved per unit co-volume of $N$ D3 branes. The information of the D1 branes is 
stored in the relation $ \tan \psi=\frac{N_{(1)}}{N}$. It can also be seen that the asymptotic value of the B field is $B^{\infty}_{23}=\tan \psi$ while $R$ is
related to the other parameters via $R^{4}=4 \pi g_{s} N \cos \psi$.

The proper decoupling limit of the above stated solution resembles the field theoretic limit of the noncommutative open string\cite{Seiberg:1999vs}\cite{Bilal:2000bk}, 
for which the asymptotic value of $B_{23}$ has to be scaled to infinity in a certain way. 
\begin{align}
 \tan \psi \rightarrow \frac{\theta}{\alpha^{\prime}}\;\;;\;\;x_{(2,3)} \rightarrow \frac{\alpha^{\prime}}{\theta} x_{(2,3)}
 \;\;;\;\;r \rightarrow \alpha^{\prime} R^{2} u\;\;;\;\;g_{s}\rightarrow \frac{\alpha^{\prime}}
 {\theta} \hat{g}\;\;;\;\;\alpha^{\prime} \rightarrow 0
 \end{align}
With the above scaling and keeping $x_{(2,3)}$, $u$, $\hat{g}$, $\theta$ fixed the resulting metric and field configurations are given by
\begin{align}
\label{dual}
&\;\;\;ds^{2}_{str}=\alpha^{\prime}\sqrt{\lambda}u^{2}\;[-dx^{2}_{0}+dx^{2}_{1}+h(dx^{2}_{2}+dx^{2}_{3})\;]+\alpha^{\prime}\sqrt{\lambda}\frac{du^{2}}{u^{2}}+
\alpha^{\prime}\sqrt{\lambda}d\Omega^{2}_{5}\nonumber\\
&\;\;\;\;h=\frac{1}{1+\lambda \theta^{2} u^{4}}\;\;\;;\;\;\;B_{23}=\frac{\alpha^{\prime}\lambda \theta u^{4}}{1+\lambda \theta^{2}u^{4}}\;\;\;;\;\;\;e^{2 \phi}=
\hat{g}h\;\;\;;\;\;\;\lambda\equiv R^{4}=4 \pi \hat{g} N
\end{align}
The above is the holographic dual to NCYM with gauge group SU(N) and Yang Mills coupling constant $g_{NCYM}=\sqrt{4\pi\hat{g}}$ which captures the dynamics of NCYM . As the holographic
direction $u$ tends to infinity, 
the first two directions $x_{0},x_{1}$ scale as $u^{2}$ and the $x_{2},x_{3}$ directions develop a $\frac{1}{u^{2}}$ dependence in the metric.
Due to the noncommutativity the symmetry group of the theory becomes $SO(1,1)\otimes SO(2)$ as can be clearly seen from the isometry of \eqref{dual}\cite{Hashimoto:1999ut}. The gravity dual to NCYM at 
finite temperature T is found from the near
horizon limit of the black D1-D3 solution and reads
\begin{align}
 \label{thdual}
  \;\;ds^{2}_{str}=\alpha^{\prime}\sqrt{\lambda}u^{2}\;[-\Big( 1-\frac{\pi^{4}T^{4}}{u^{4}}\Big)dx^{2}_{0}&+dx^{2}_{1}+\frac{1}{1+\lambda \theta^{2} u^{4}}(dx^{2}_{2}
   +dx^{2}_{3})\;]+\frac{\alpha^{\prime}\sqrt{\lambda}}{\Big( 1-\frac{\pi^{4}T^{4}}{u^{4}}\Big)}\frac{du^{2}}{u^{2}}\nonumber\\
 &B_{23}=\frac{\alpha^{\prime}\lambda \theta u^{4}}{1+\lambda \theta^{2}u^{4}}
\end{align}

The most rigorous approach to study Schwinger effect in the holographic setting is to find the expectation value of circular Wilson loop and relate it to pair production rate as in \cite{Semenoff:2011ng}. However one may 
think of the vacuum to be made of  $q\bar{q}$ pairs bound under an attractive potential and study how an external electric field modifies this potential. This is the
essence of potential analysis which was first put forward in \cite{Sato:2013iua}. To compute the inter-quark potential, one needs to look at expectation value of
rectangular Wilson Loop when the loop contour is regarded as trajectory of particles under consideration in the $x_{0}-x_{3}$ plane where $x_{3}$ is the 
direction of $q\bar{q}$ orientation. As pointed out in \cite{Semenoff:2011ng}\cite{Fischler:2014ama} one places a probe D brane at a finite position instead of the boundary to get a W boson 
of finite mass.\footnote{A string with Dirichlet conditions at both ends has the following canonical Hamiltonian \cite{Blumenhagen:2013fgp}, where first term 
indicates the potential energy of the stretched string and is the analogue of mass created due to symmetry breaking.
\begin{equation}
\label{DD}
H=\frac{{(q^{\mu}_{a}-q^{\mu}_{b})}^{2}}{4\pi \alpha^{\prime}}+\sum_{n\neq 0} \alpha_{(-n)}\alpha_{(n)}\nonumber
\end{equation}} As per the holographic procedure
the VeV of the Wilson Loop of a gauge theory is given by the partition function of a fundamental string in the background of the holographic dual with the ends of the 
string anchored on the probe D brane along the contour of the Wilson Loop $\mathcal{C}$ \eqref{wloopqft} i.e.
\begin{align}
 \label{wloopstring}
 \expval{W[\mathcal{C}]}=\frac{1}{\text{Vol}}\int_{\partial X=\mathcal{C}} \mathcal{D}X \mathcal{D}h_{ab}\;e^{-S[X,h]}
\end{align}
Where $S[X,h]$ indicates the action of the fundamental string\footnote{The boundary conditions is given by
trajectory of the string at the probe brane , $\partial X=\mathcal{C}$. Stated more explicitly $\mathcal{D}X=\mathcal{D}\xi$ where $X^{\mu}(s,t)=c^{\mu}(t)+
\xi^{\mu}(t,s)$. The functions $\xi^{\mu}(t,s)$ vanishes at the boundary which is given by a specific value $s=s_{0}$, $c^{\mu}(t)$ being parametric representation 
of the contour $\mathcal{C}$ of the Wilson Loop.} which has been Wick rotated to euclidean signature, as is usually done in string theory \cite{Blumenhagen:2013fgp}. In the classical limit which is realized
when the string length $\alpha^{\prime}$ is small (or the 't Hooft coupling 
is big ) the above expression is dominated by the on shell value of the Polyakov/Nambu Goto action. Thus the prescription of computing Wilson loops is reduced to
computing the area of the world sheet  of the fundamental string which end on the specified profile at the probe D brane \cite{Drukker:1999zq}
\cite{Maldacena:1998im}\cite{Rey:1998ik}, situated at a finite radial position in the dual geometry for the present case.

\subsection{Potential Analysis at Zero Temperature}

\begin{figure}[tbp]
\centering 
\includegraphics[width=1.00\textwidth,height=0.60\textwidth,origin=c,angle=0]{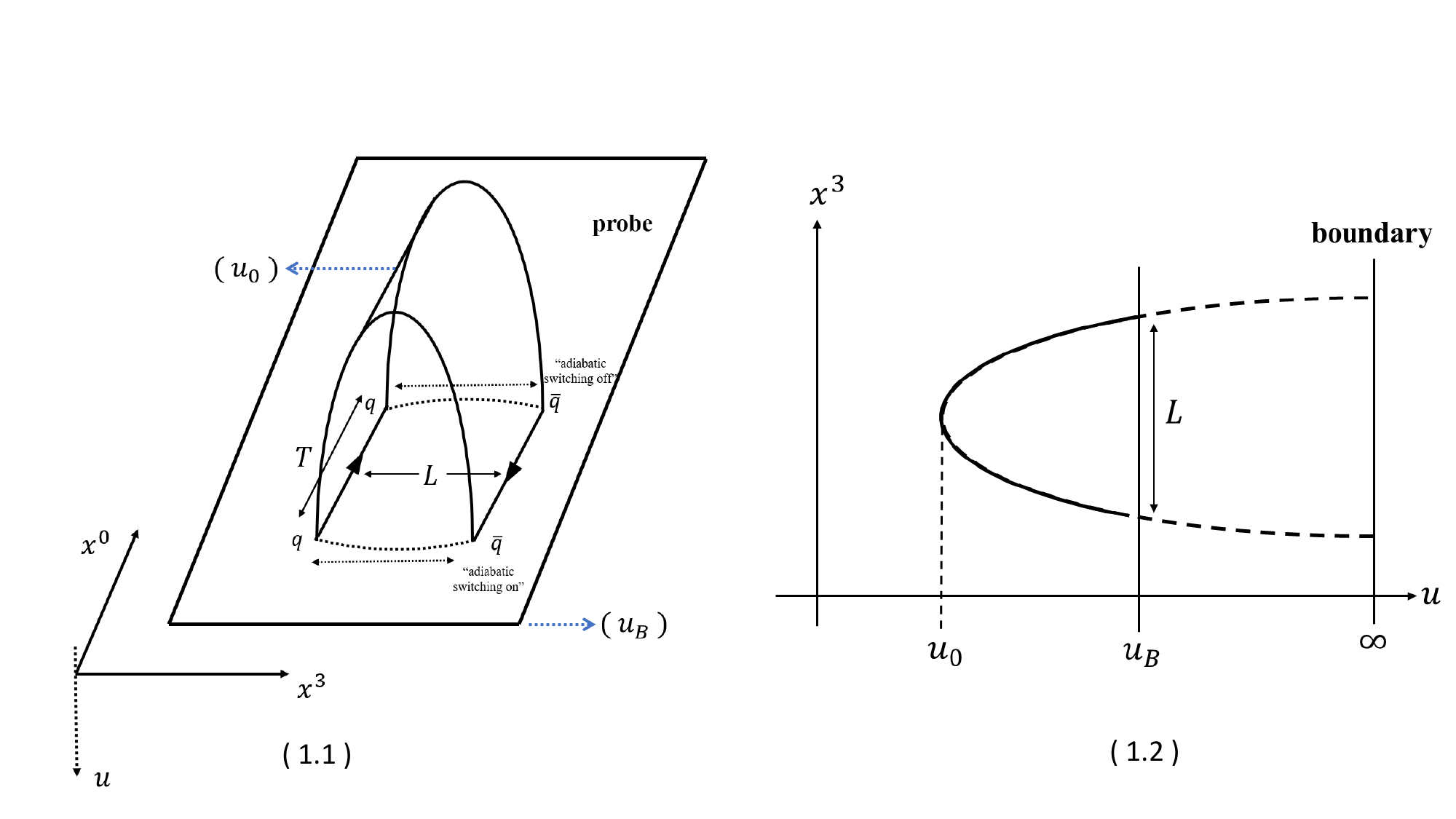}
\caption{This figure illustrates the setup used. The probe brane is placed at a finite position ($u_{B}$) on the holographic direction as in (1.2). On the probe brane
 the placement of the Wilson loop is shown in (1.1), arrows indicating the contour of the loop (not the propagation of the string). For adiabatic interactions one can neglect the effects of the
 dotted lines and the string profile becomes static. }
\label{fig1}
\end{figure}

In this section we will study the properties of the modified potential of NCYM in presence on a constant external electric field for quark/anti-quark pairs along one of 
the noncommutative directions ($x_{3}$) . The appropriate holographic dual as pointed above is given by \eqref{dual}. To calculate the extremal area of the string configuration ending of the probe brane
we take the string world 
sheet to be parametrized by $s^{a} \equiv (s,t)$. The Nambu Goto action reads
\begin{align}
 \label{NG}
 & \mathcal{S}_{NG}=\frac{1}{2 \pi \alpha^{\prime}}\int dt ds \sqrt{det\;G^{(in)}_{ab}}\nonumber\\
 & \;\;\;G^{(in)}_{ab} \equiv G_{\mu\nu} \frac{\partial x^{\mu}}{\partial s^{a}} \frac{\partial x^{\nu}}{\partial s^{b}}
\end{align}
In the above $G^{(in)}_{ab}$ is the induced metric on the worldsheet and $G_{\mu\nu}$ is the metric of the target spacetime/holographic dual\eqref{dual}\eqref{thdual}.
The above action exhibits two diffeomorphism symmetries with the help of which one can set two of the embedding functions to arbitrary values provided that the
resulting profile matches with the contour of our choosing on the probe brane. One usually chooses the so called static gauge for which the profile reads
\footnote{Strictly speaking the embedding function which extremizes the Nambu Goto function may not respect such a gauge choice globally and may lead us to a local
minimum of the Nambu Goto action. One can hope the the results found will converge to the true value if perturbations are added.}
\begin{align}
 \label{staticgauge}
 x^{0}(s,t)=t\;\;;\;\;x^{3}(s,t)=s\;\;;\;\;u(s,t)=u(s)\;\;;\;\;x^{1,2}=0\;\;;\;\;\Theta^{i}(s,t)=\text{constant}
\end{align}
In the above the extremization of the Nambu Goto action is given by the functional form of $u(s)$ .The $\Theta^{i}$ are the co-ordinates of $S^{5}$. For present
purposes $x_{3} \equiv s $ is assumed to to range between $[-L,L]$, when $2L$ indicates the inter-quark separation on the probe brane with the boundary 
condition $u(\pm L)=u_{B}$ where $u_{B}$ indicates the position of the probe brane along the holographic direction. Again the temporal direction is assumed to 
range from $[-\mathcal{T},\mathcal{T}]$ with the further assumption that $\mathcal{T}\gg L$, meaning time scale of
the problem (within which the quarks/anti-quark pairs remain seperated) to be much larger
than the length scale. This is because the rectangular Wilson Loop gives the inter-quark potential  when
one assumes the interaction between dipole is adiabatically switched on and off as illustrated in Figure \ref{fig1}.\footnote{Intuitively one thinks, two quarks to be separated at
distant past and then reunite at distant future. 
Thus the worldline becomes a rectangular closed contour whose area can be identified with the inter-quark potential $\expval{W[\mathcal{C}]}\sim e^{-\mathcal{T}U(L)}$.}

Before proceeding further let us address the issue of the B field. For noncommutative gauge theories the  B field is excited \eqref{dual}\eqref{thdual} and is 
present in the string action via the Wess-Zumino term $\int dt ds B_{\mu\nu}\partial_{t}x_{\mu}\partial_{s}x_{\nu}$ along with the usual Polyakov/Nambu Goto part.
However the gauge choice given above \eqref{staticgauge} canecls the contribution of the Wess-Zumino part of the action. It is possible to consider 
the $q\bar{q}$ pairs at a velocity in the $x_{2}$ direction and take into account the contribution of the $B_{\mu\nu}$ term as in \cite{Dhar:2000nj}. However in the
present case where the virtual particles in vacuum are modeled as $q\bar{q}$ dipoles, such a configuration seems hardly sensible.

As per the above gauge choice the induced metric/line element on the world sheet reads.
\begin{align}
 \label{Gin}
 G^{(in)}_{ab}ds^{a}ds^{b}=-\alpha^{\prime}\sqrt{\lambda}u^{2}dt^{2}+\frac{\alpha^{\prime}\sqrt{\lambda}}{u^{2}}\bigg[\;\big(\frac{du}{ds}\big)^{2}+
 \frac{u^{4}}{1+\lambda \theta^{2} u^{4}}\;\bigg]ds^{2}
\end{align}
Using the above form of the induced metric in the action \eqref{NG}, we get
\begin{align}
\label{mNG}
 \mathcal{S}_{NG}=\frac{\mathcal{T}}{2 \pi \alpha^{\prime}} \int_{-L}^{L} ds \; \sqrt{\big(\frac{du}{ds}\big)^{2}+\frac{u^{4}}{1+\lambda \theta^{2} u^{4}}}
\end{align}
Extremization of the above Lagrangian is equivalent to solving the Euler Lagrange equation with the effective Lagrangian $\mathcal{L}_{NG}=\sqrt{\big(\frac{du}{ds}
\big)^{2}+\frac{u^{4}}{1+\lambda \theta^{2} u^{4}}}$ when $u=u(s)$ along with the the boundary condition $u(s \equiv x_{3}=\pm L)=u_{B}$ as the contour profile is 
already taken into account by the gauge choice. One can indeed solve the relevant problem and find the explicit form of $u(s \equiv x_{3})$. However since the 
Lagrangian in \eqref{mNG} does not explicitly depend on the parameter $s$\footnote{In principle one consider a profile where $u=u(s,t)$ as the two diffeomorphism
symmetries are exhausted by \eqref{staticgauge}. However such a choice would reflect non-adiabatic interactions i.e. the 
interpretation of Wilson loops of measuring the interaction potential of $q\bar{q}$ pairs at rest as in Figure \ref{fig1} would
be invalid.}, by Noether's theorem there exists a conserved quantity ($Q$) for the solution.
\begin{align}
 \label{charge}
Q \equiv -\frac{du}{ds}\frac{\partial \mathcal{L}_{NG}}{\partial \big(\frac{du}{ds}\big)}+\mathcal{L}_{NG}=
\frac{ u^{4}h(u) }{\sqrt{\big( \frac{du}{ds} \big)^{2}+u^{4} h(u)}}\;\;\;\;\;;\;\;\;\;\;h(u)=\frac{1}{1+\lambda \theta^{2}u^{4}}
\end{align}
As indicated in \cite{Rey:1998ik} the fundamental string is assumed to be carrying charges at its two ends and are otherwise symmetric about its origin. Thus one can 
choose $u(s)$ to be an even function of $s \equiv x_{3} $. In the present case this means the $x_{3}$ direction of NCYM is symmetric around its origin which holds true
in-spite of its noncommutative nature. From the form of \eqref{charge} the solution of $\frac{du}{ds}$ involves both positive and negative signs. Thus there exist a
value of the parameter $s=s_{0}$ for which $\frac{du}{ds}(s_{0})=0$. This is the turning point of the string profile as indicated in Figure \ref{fig1}. Simplifying
\eqref{charge} and introducing a rescaled holographic co-ordinate $y=\frac{u}{u_{0}}$ one obtains the following differential equation 
\begin{align}
 \label{diffeq}
 \frac{d}{ds}\big(\frac{u}{u_{0}}\big) \equiv \frac{dy}{dx_{3}}=\frac{u_{0}y^{2}\sqrt{y^{4}-1}}{1+\lambda \theta^{2}u_{0}^{4}y^{4}}
\end{align}
In the above $u_{0}$ indicates the value of $u(s)$ at $s=s_{0}$ and the gauge choice $x_{3}(s,t)=s$ has been used. The above equation is obtained from evaluating
the l.h.s of \eqref{charge} at the turning point. From the equation obtained one can estimate the separation length ($2L$) of the $q\bar{q}$ dipole by integration
both sides 
\begin{align}
 \label{dipoleintegral}
 L \equiv \int _{0}^{L} dx_{3}= \int_{1}^{\frac{u_{B}}{u_{0}}}  \frac{dy}{u_{0} y^{2} \sqrt{y^{4}-1}}+\int_{1}^{\frac{u_{B}}{u_{0}}}  
 \frac{\lambda \theta^{2}u_{0}^{3}y^{2}}{\sqrt{y^{4}-1}} \; dy 
\end{align}
It is worthwhile to point out that if one tries to take $u_{B}\rightarrow \infty$  in \eqref{dipoleintegral} the dipole length diverges due to the second integral
(which is absent in the commutative counterpart where $\theta=0$). However unlike the generic quark/anti-quark potential calculation where a divergence is attributed 
to the self energy of infinitely massive quarks, the present situation cannot be remedied by such arguments. The fact that the holographic dual of a NCYM does not live
at radial infinity has been reported in \cite{Maldacena:1999mh} where it has been shown a slight perturbation on the string profile at infinity destabilizes it 
completely. An alternative has been advocated in \cite{Dhar:2000nj} where the string profile is allowed to have a velocity in the $x_{2}$ and the $B_{\mu\nu}$ term in
the string action contributes to the inter-quark length unlike the present case. It can be shown that for a certain velocity of the $q\bar{q}$ pair in the transverse 
direction the dipole can be consistently taken to radial infinity. As indicated before we avoid such a configuration for the present case.

The mass of the fundamental matter ($q\bar{q}$ pairs)  is given by the self energy of a stretched string from the probe to the interior \cite{Kinar:1998vq}. For 
determining the same the relevant gauge is  $x_{0}=t$, $u=s$, $x_{3}=$constant. Thus the mass is given by
\begin{align}
 m=\frac{1}{2\pi \alpha^{\prime}}\int_{0}^{u_{B}} du \sqrt{\alpha^{\prime}\sqrt{\lambda}u^{2}\cdot \frac{\alpha^{\prime}\sqrt{\lambda}}{u^{2}}}
 =\frac{\sqrt{\lambda}}{2\pi}u_{B}
\end{align}
The dipole separation length of the test particles \eqref{dipoleintegral} can be analytically integrated and in terms of the
parameter $a=\frac{u_{0}}{u_{B}}$ one has 
\begin{align}
 \label{dipolelength1}
&L=\frac{\sqrt{\lambda}}{2\pi m}\bigg[\frac{\sqrt{\pi}\;\Gamma(3/4)}{a\;\Gamma(1/4)}-
\frac{a^{2}}{3} \mathstrut_2 F_1\bigg(\frac{1}{2},\frac{3}{4},\frac{7}{4},a^{4}
\bigg)\bigg]\nonumber\\
&\;\;\;\;\;\;\;+\frac{8\pi^{3}m^{3}\theta^{2}}{\sqrt{\lambda}}a^{3}\bigg[-\frac{\sqrt{\pi}\;\Gamma(3/4)}{\Gamma(1/4)}
+a\cdot\mathstrut_2 F_1\bigg(-\frac{1}{4},\frac{1}{2},\frac{3}{4},a^{4}\bigg)\bigg]
\end{align}
\begin{figure}[tbp]
\centering 
\includegraphics[width=.60\textwidth,origin=c,angle=0]{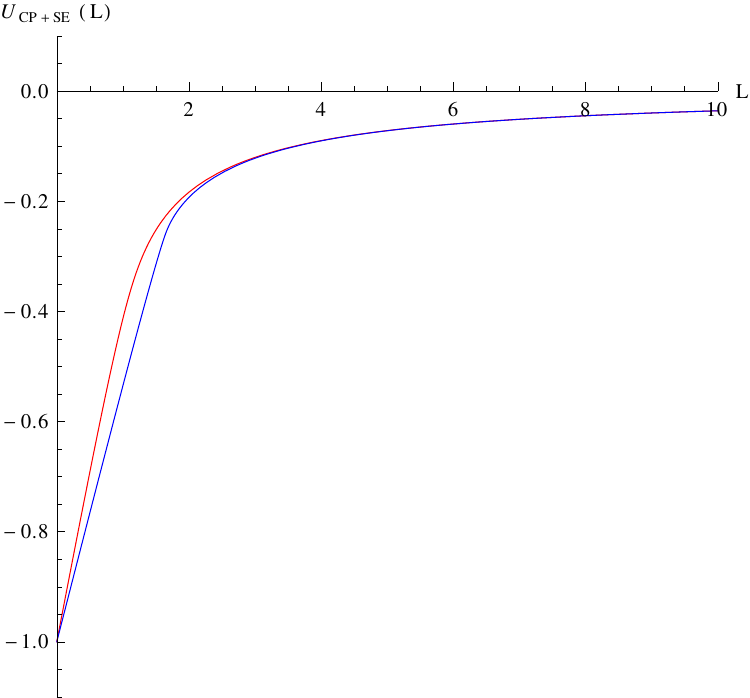}
\caption{This is a parametric plot of $U_{CP+SE}$ v/s L. The plot indicates the profile of the potential (with static energy subtracted) to be coulombic at large
distances. At small distance the profile exhibits a linear dependence. However noncommutativity has a tendency to increase the linear effect. The values used are
$m=1$, $\lambda=4 \pi^{2}$ . The \textcolor{red}{red} line stands for the value \textcolor{red}{$\theta=0.2$} while the \textcolor{blue}{blue} line indicates
\textcolor{blue}{$\theta=0.3$}.}
\label{fig2}
\end{figure}
The sum of the potential and the static energy of the $q\bar{q}$ pairs is given by the (time averaged) on-shell value of the Nambu Goto action  which by the virtue of
\eqref{mNG},\eqref{dipoleintegral} is 
\begin{align}
 \label{Potential}
 U_{CP+SE}=&\frac{\sqrt{\lambda}}{2\pi}a u_{B}\int_{1}^{1/a}\;dy\;\frac{y^{2}}{\sqrt{y^{4}-1}}\;\sqrt{1+\lambda \theta^{2} a^{4} u_{B}^{4}}\nonumber\\
 =& m\sqrt{1+\frac{16\pi^{4}m^{4}\theta^{2}a^{4}}{\lambda}} \bigg[- a \frac{\sqrt{\pi}\;\Gamma(3/4)}{\Gamma(1/4)} +
 \mathstrut_2 F_1\bigg(-\frac{1}{4},\frac{1}{2},
 \frac{3}{4},a^{4}\bigg) \bigg]
\end{align}
From the above one can look at the limit when $L \rightarrow \infty  $ which is the same as taking the limit $a \rightarrow 0$. This fact is evident from the above
expression, see Figure \ref{fig3}. This is the situation when  the string stretches to the interior i.e. $u_{0} \rightarrow 0$ . Using the relation $\mathstrut_2 F_1 (-\frac{1}{4},\frac{1}{2},
\frac{3}{4},0)=1$ the leading dependence of the inter-quark length \eqref{dipolelength1} is given by $L=\frac{1}{2m}\sqrt{\frac{\lambda}{\pi}} \frac{\Gamma(3/4)}{a
\;\Gamma(1/4)}$, and the inter-quark potential \eqref{Potential} becomes
\begin{align}
U_{CP+SE}(a \rightarrow 0)\approx m-am\frac{\sqrt{\pi}\;\Gamma(3/4)}{\Gamma(1/4)}=m-
\frac{\sqrt{\lambda}}{2}\bigg(\frac{\Gamma(3/4)}{\Gamma(1/4)}\bigg)^{2}\frac{1}{L}
\end{align}
\begin{figure}[tbp]
\centering 
\includegraphics[width=.60\textwidth,origin=c,angle=0]{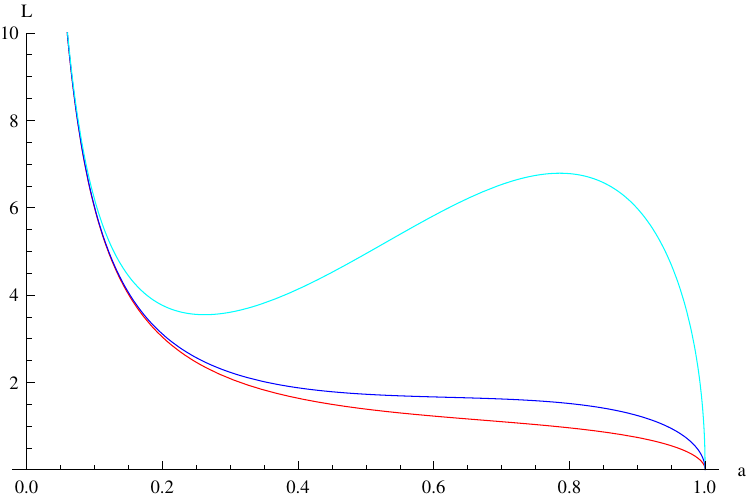}
\caption{ This is a parametric plot of $L(a)$ v/s $a$. The values chosen are $m=1$, $\lambda=4 \pi^{2}$.The \textcolor{red}{red} and \textcolor{blue}{blue} lines are 
plots for \textcolor{red}{$\theta=0.2$} and \textcolor{blue}{$\theta=0.3$} respectively, the \textcolor{cyan}{cyan} line stands for \textcolor{cyan}{$\theta=0.75$}.
Note that for the later case the plot becomes degenerate of intermediate values of the parameter $a$, though the predictions for extreme values of $L$ remain true.
We will stick to the first two values for our numerical computations.}
\label{fig3}
\end{figure}
Thus the usual Coulomb law is recovered at large distances . However for arbitrary separation there is a rapid modification from the Coulomb law. This can be
attributed due to two reasons. Firstly as is evident from \eqref{dipolelength1} and \eqref{Potential} for arbitrary values of the parameter a, the noncommutative 
effects creep in which breaks the conformal symmetry and hence coulombic dependence. Secondly as found in \cite{Sato:2013iua}, even for a commutative theory
the potential profile is altered from and is finite even at short distances. This is because in presence of a mass term the theory is not conformal anymore as can be
understood from the presence of \eqref{sw} which is coupled to the usual SU(N) action. To get a better view of the same we look at a \textrightarrow $1$ limit of
\eqref{Potential} and \eqref{dipolelength1}. It is evident from the integrals that both of them vanishes in the above said limit. Looking at the limiting values 
one has, 
\begin{align}
 \label{Ulimiting}
 U_{CP+SE} (a\rightarrow1^{-}) \approx \bigg[&\frac{\frac{m}{\lambda}}{\sqrt{1+\frac{16\pi^{4}m^{4}\theta^{2}a^{4}}{\lambda}}}\bigg(-
 \frac{16\pi^{4}m^{4}\theta^{2}a^{4}
 +\lambda}{a\sqrt{1-a^{4}}}-\frac{\sqrt{\pi}(48\pi^{4}m^{4}\theta^{2}a^{4}+\lambda)\Gamma(\frac{3}{4})}{\Gamma(\frac{1}{4})}\;\nonumber\\
&+\frac{(48\pi^{4}m^{4}\theta^{2}a^{4}+\lambda)\mathstrut_2 F_1\big(-\frac{1}{4},\frac{1}{2},\frac{3}{4},a^{4}\big)}{a}\bigg)\bigg]_{a\rightarrow 1^{-}}\cdot(1-
a)
\end{align}
In the above the $\mathcal{O}(1-a)^{2}$ terms have been neglected. Similarly for the inter-quark distance one has,
\begin{align}
 \label{dipolelength1limiting}
 L (a\rightarrow1^{-})\approx \;& (1-a)\cdot\frac{\sqrt{\lambda}}{2\pi m}\bigg[-\frac{a}{\sqrt{1-a^{4}}}+\frac{a}{3}\mathstrut_2 F_1\big(\frac{1}{2},\frac{3}{4},
 \frac{7}{4},a^{4}\big)-\frac{\sqrt{\pi}}{a^{2}}\frac{\Gamma(\frac{3}{4})}{\Gamma(\frac{1}{4})}\bigg]_{a\rightarrow 1^{-}}\nonumber\\
 &+(1-a)\cdot\bigg[-\frac{8\pi^{3}m^{3}\theta^{2}a^{3}}{\sqrt{\lambda}\sqrt{1-a^{4}}}+\frac{16\pi^{3}m^{3}\theta^{2}a^{3}}{\sqrt{\lambda}}\mathstrut_2 F_1\big(
 \frac{-1}{\;\;4},\frac{1}{2},\frac{3}{4},a^{4}\big)\bigg]_{a\rightarrow1^{-}}
\end{align}
Comparing \eqref{Ulimiting} and \eqref{dipolelength1limiting} and using the identity $\mathstrut_2 F_1\big(a,b,c,1\big)=\frac{\Gamma(c)\Gamma(c-a-b)}{\Gamma(c-a)
\Gamma(c-b)}$ , it can be easily seen that for short inter-quark separation .
\begin{align}
 \label{shortdistance}
 U_{CP+SE} (a\rightarrow1^{-}) \approx \frac{2\pi m^{2}}{\sqrt{\lambda}}\frac{1}{\sqrt{1+\frac{16\pi^{4}m^{4}\theta^{2}}{\lambda}}} L (a\rightarrow1^{-})
\end{align}
Thus increment in the value of noncommutative parameter $\theta$ results into decrement of the the slope of the potential curve at small separation. This is 
because of repulsive forces due to noncommutativity, signaling the force needed to detach the noncommutative $q\bar{q}$ pair should be smaller than its
commutative counterpart.

Till now we have calculated the inter-quark potential. However in presence of and external electric field the charged $q\bar{q}$ pairs develop an electrostatic 
potential of their own. The total potential is given by the sum of the two. Defining the effective potential to be ,
\begin{align}
\label{total}
 U_{effective}\;(L)=U_{CP+SE}\;(L)-E\cdot L
\end{align}
It can be guessed from \eqref{shortdistance} that in the presence of an external electric field of strength,
\begin{align}
\label{critical}
\mathcal{E}_{T}= \frac{2\pi m^{2}}{\sqrt{\lambda}}\frac{1}{\sqrt{1+\frac{16\pi^{4}m^{4}\theta^{2}}{\lambda}}}
\end{align}
the $q\bar{q}$ pair overcomes linear barrier of the potential profile in Figure \ref{fig2}. However it is still to be seen whether the above mentioned electric field
is enough to get out of the tunneling phase for all values of inter-quark separation $L$. One has from \eqref{dipoleintegral} and \eqref{Potential},
\begin{align}
 \label{effective}
 &U_{effective}\;(L(u_{0}))=\Big(1-r \Big)\mathcal{E}_{T}L(u_{0})+G\;(L(u_{0}))\nonumber\\
 &G\;(L(u_{0}))=\frac{\sqrt{\lambda}}{2 \pi}\int_{u_{0}}^{u_{B}}du\bigg[\frac{u^{2}\sqrt{1+\lambda \theta^{2}u_{0}^{4}}}{\sqrt{u^{4}-u_{0}^{4}}}
 -\frac{u_{B}^{2}u_{0}^{2}}{\sqrt{1+\lambda \theta^{2}u_{B}^{4}}}\frac{1+\lambda \theta^{2}u^{4}}{u^{2}\sqrt{u^{4}-u_{0}^{4}}}\bigg]\nonumber\\
&L(u_{0})=\int_{u_{0}}^{u_{B}}du\frac{u_{0}^{2}}{u^{2}}\frac{\big(1+\lambda \theta^{2}u^{4}\big)}{\sqrt{u^{4}-u_{0}^{4}}}
\end{align}
\begin{figure}[tbp]
\centering 
\begin{subfigure}[b]{0.50\textwidth}
                \centering
                \includegraphics[width=\textwidth,origin=c,angle=0]{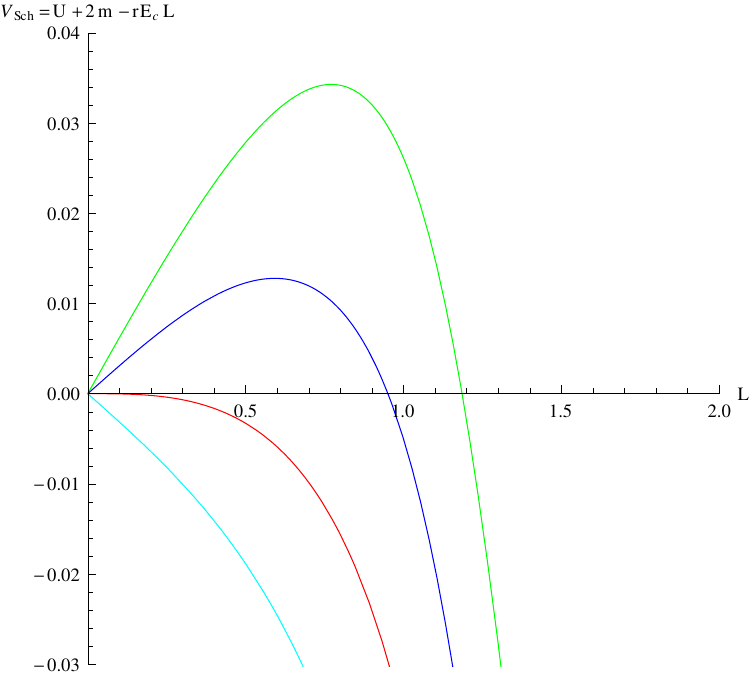}
        \subcaption{$\theta=0.2$}
        \end{subfigure}%
        \begin{subfigure}[b]{0.50\textwidth}
                \centering
                \includegraphics[width=\textwidth,origin=c,angle=0]{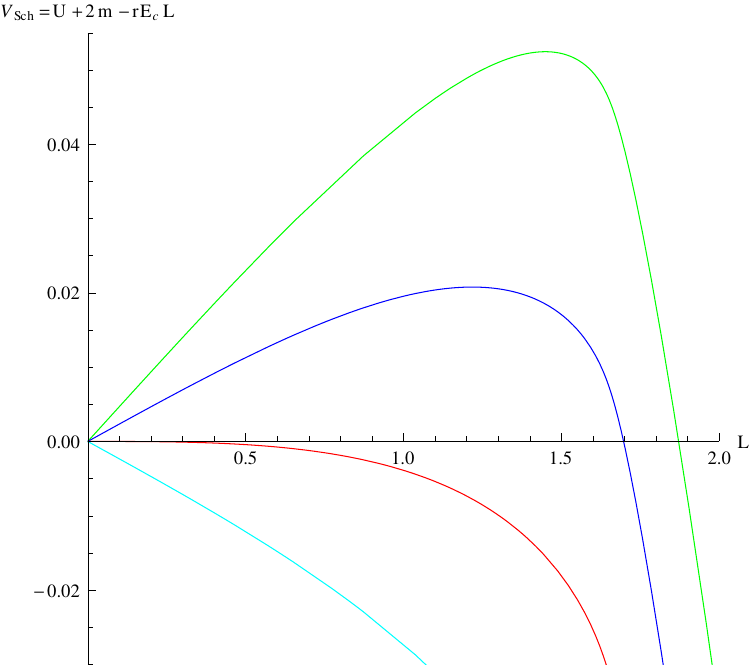}
                \subcaption{$\theta=0.3$}
        \end{subfigure}%
\caption{The plot indicates the effective potential (in presence of external electric field) v/s the inter-quark separation. The values used are $m=1$ and
$\lambda=4\pi^{2}$. The \textcolor{green}{green} line indicates \textcolor{green}{r=0.9}, \textcolor{blue}{blue} line for \textcolor{blue}{r=0.95}. The parameter
r is the ratio of the applied
field to it's threshold value. Note that the maximal height of the potential barrier decreases as noncommutativity increases. The 
\textcolor{red}{red} line which exhibits the threshold behavior stands for \textcolor{red}{r=1.0} and \textcolor{cyan}{cyan} for \textcolor{cyan}{r=1.05} shows
catastrophic decay of vacuum.}
\label{fig4}
\end{figure}
In the above we have reinstated the turning point $u_{0}$ and have introduced the ratio $r=\frac{E}{\mathcal{E}_{T}}$ for simplicity where $\mathcal{E}_{T} $, the 
threshold electric field is given by \eqref{critical}. It is apparent from \eqref{effective} , at the value $r=1$ (applied field being of threshold value) the first 
term vanishes. Thus the potential profile is governed fully by the second term $G\;(L(u_{0}))$ , and will cease to put up a tunneling barrier if the function is
monotonically decreasing and is vanishing at the origin. At $L=0$ which is realized if $u_{0}=u_{B}$ it is evident from \eqref{effective} that $G\;(L=0)=0$. Indeed 
this is the case for $U_{effective}(L=0)$ too. 
\begin{align}
\label{slope}
 \frac{d}{dL}U_{effective}\;(L)=(1-r)\mathcal{E}_{T}+\frac{d}{dL}G\;(L)=(1-r)\mathcal{E}_{T}+\frac{du_{0}}{dL}\frac{dG\;(u_{0})}{du_{0}}
\end{align}
Moreover one can show from \eqref{effective} that,
\begin{align}
 \label{Lprime}
 \frac{dL (u_{0})}{du_{0}}&=-\frac{\big(1+\lambda \theta^{2}u_{0}^{4}\big)}{\sqrt{\big(u_{0}+\varepsilon \big)^{4}-u_{0}^{4}}}+\int_{u_{0}}^{u_{B}}du
 \bigg[\frac{2u_{0}\big(1+\lambda \theta^{2}u^{4}\big)}{u^{2}\sqrt{u^{4}-u_{0}^{4}}}+\frac{2u_{0}^{5}\big(1+\lambda \theta^{2}u^{4}\big)}{
 u^{2}\big(\sqrt{u^{4}-u_{0}^{4}}\big)^{3}}\bigg]\nonumber\\
 &=\bigg[-\frac{\big(1+\lambda \theta^{2} u_{0}^{4}\big)}{\sqrt{\big(u_{0}+\varepsilon \big)^{4}-u_{0}^{4}}}+
 2\int_{u_{0}}^{u_{B}}du\frac{\big(1+\lambda \theta^{2}u^{4}\big)}{\big(\sqrt{u^{4}-u_{0}^{4}}\big)^{3}}u_{0}u^{2}\bigg]
\end{align}
The first term comes when the differential operator acts on the lower limit of integration in \eqref{effective}. A regulator $\varepsilon$ (whose physical meaning is
rather vague) has been put in the expression as the first term is actually divergent. By a similar procedure one has .
\begin{align}
 \label{Gprime}
 \frac{dG (u_{0})}{du_{0}}=\frac{\sqrt{\lambda}}{2 \pi}&\bigg[-\frac{\big(1+\lambda \theta^{2}u_{0}^{4}\big)}{\sqrt{\big(u_{0}+\varepsilon \big)^{4}-u_{0}^{4}}}
 +2\int_{u_{0}}^{u_{B}}
 u_{0}u^{2}\frac{\big(1+\lambda \theta^{2}u^{4}\big)}{\big(\sqrt{u^{4}-u_{0}^{4}}\big)^{3}}\bigg]\nonumber\\
 &\times\bigg(\frac{u_{0}^{2}\sqrt{1+\lambda \theta^{2} u_{B}^{4}}-u_{B}^{2}
 \sqrt{1+\lambda \theta^{2}u_{0}^{4}}}{\sqrt{\big(1+\lambda \theta^{2} u_{B}^{4}\big)\big(1+\lambda \theta^{2}u_{0}^{4}}\big)}\bigg)
\end{align}
Though the above two terms are actually divergent one can see from \eqref{Lprime} and \eqref{Gprime} that their ratio isn't.
\begin{align}
\label{ratio}
G\;^{\prime}(L)\equiv\frac{dG (L)}{dL}=\frac{\sqrt{\lambda}}{2 \pi}\bigg(\frac{u_{0}^{2}}{\sqrt{1+\lambda \theta^{2} u_{0}^{4}}}-
\frac{u_{B}^{2}}{\sqrt{1+\lambda \theta^{2}u_{B}^{4}}} \bigg)
\end{align}
It is easily seen that $G(L(u_{0}))$ is a monotonically decreasing function w.r.t $L(u_{0})$ i.e for $u_{B}\geq u_{0}$ . From the above it is clear that the net
potential/force due to the applied field has two components , 
\begin{itemize}
 \item[(a)] $(1-r)\mathcal{E}_{T}L(u_{0})$ : This is the part which creates the potential barrier for $r<1$ i.e the attractive force between the $q\bar{q}$ pairs in an 
 external electric field. At $r=1$ this part ceases to contribute and for $r>1$ the force corresponding to this part becomes repulsive.
 \item[(b)] $G\;(L(u_{0}))$ : This part contributes to bringing down the potential barrier. As can be seen from \eqref{ratio}, the associated force due to this is 
 repulsive for all values of $L(u_{0})$ except at the origin where it vanishes. Thus at threshold point ($r=1$) where the first component (part a) becomes irrelevant 
 the slope of the net potential for all nonzero values $L$ is negative as can be seen in Figure \ref{fig4} confirming the prediction of \eqref{critical}.
\end{itemize}

It is also interesting to see whether the effective potential admits a confining phase where the tunneling behavior is totally absent. This amounts to showing 
the existence of a intermediate value  of $u_{0}$ for $r<1$ where the total potential as in \eqref{total},\eqref{effective} vanishes. Alternatively one
can check the values of the slope of the potential at extreme points and look for a saddle point of the same. It is easy to see that at $L=0/u_{0}=u_{B}$ the slope 
\eqref{slope} is given by $(1-r)\mathcal{E}_{T}$ which is positive for $r < 1$. However at $u_{0}=0 / L\rightarrow \infty $ the slope of the potential is given by,
\begin{align}
\frac{d}{dL}U_{effective}(L\rightarrow \infty)=-r\frac{\sqrt{\lambda}}{2\pi}
\frac{u_{B}^{2}}{\sqrt{1+\lambda \theta^{2} u_{B}^{4}}}
\end{align}
Thus we see that the slope of the potential curve is negative (force between $q\bar{q}$ pairs is repulsive) at large distances for all values of the applied electric 
field unlike the situations in \cite{Sato:2013dwa} \cite{Ghodrati:2015rta}, indicating the present case of not being confining. It is also clear from \eqref{slope} and \eqref{ratio} that the
maximal potential barrier is encountered when,
\begin{align}
\label{nceffect}
 a^{4}\equiv \big(\frac{u_{0}}{u_{B}}\big)^{4}=\frac{r^{2}}{1+\frac{16\pi^{4}m^{4}\theta^{2}}{\lambda}\big(1-r^{2}\big)}
\end{align}
This is the point when the effective $q\bar{q}$ potential becomes repulsive rather than being attractive.

\subsection{Potential Analysis at Finite Temperature}
The finite temperature case closely resembles the above calculation. For the present situation the gravity dual is given by \eqref{thdual}. The thermal mass of
fundamental $q\bar{q}$ pairs is given by
\begin{align}
 \label{thmass}
 m(T)&=\frac{1}{2\pi \alpha^{\prime}} \int_{\pi T}^{u_{B}} du \sqrt{\alpha^{\prime}\sqrt{\lambda}u^{2}\Big(1-\frac{\pi^{4}T^{4}}{u^{4}}\Big)\cdot
 \frac{\alpha^{\prime}\sqrt{\lambda}}{u^{2}\Big(1-\frac{\pi^{4}T^{4}}{u^{4}}\Big)} }\nonumber\\
 &=\frac{\sqrt{\lambda}}{2\pi}\big(u_{B}-\pi T)=m(T=0)-\frac{\sqrt{\lambda}}{2}T
\end{align}
Very much like the  above analysis the Nambu Goto action in static gauge reduces to
\begin{align}
 \label{thNG}
 S=\mathcal{T}\frac{\sqrt{\lambda}}{2\pi}\int ds \sqrt{\big(\frac{du}{ds}\big)^{2}+\frac{u^{4}-\pi^{4}T^{4}}{1+\lambda \theta^{2}u^{4}}}
\end{align}
Quite similar to the previous case the conserved quatity arising from the Lagrangian \eqref{thNG} is 
\begin{align}
 \label{thnoether}
 \frac{1}{1+\lambda \theta^{2}u^{4}}\cdot\frac{u^{4}-\pi^{4}T^{4}}{\sqrt{\big(\frac{du}{ds}\big)^{2}+\frac{u^{4}-\pi^{4}T^{4}}{1+\lambda \theta^{2}u^{4}}}}=\text{conserved}
\end{align}
Demanding the profile admits a zero slope at the turning point $u_{0}$ (the turning point should be greater than the horizon radius i.e. $u_{0}\geq \pi T$), we have
\begin{align}
 \label{thdiffeq}
 \frac{du}{ds}=\frac{\sqrt{\big(u^{4}-\pi^{4}T^{4}\big)\big(u^{4}-u_{0}^{4}\big)\big(1+\lambda \theta^{2}\pi^{4}T^{4}\big)}}{\sqrt{
 \big(u_{0}^{4}-\pi^{4}T^{4}\big)}\big(1+\lambda \theta^{2}u^{4}\big)}
\end{align}
From the above equation the separation length between test particles can be integrated out to be 
\begin{align}
 \label{thscreening}
 L_{T}(a)=\frac{1}{a}\frac{\sqrt{\lambda}}{2\pi m}\int_{1}^{\frac{1}{a}} dy \frac{\sqrt{1-\frac{\lambda^{2}T^{4}}{16m^{4}a^{4}}}\big(1+\frac{16 \pi^{4}m^{4}\theta^{2}
 }{\lambda}y^{4}a^{4}\big)}{\sqrt{\big(y^{4}-1\big)\big(y^{4}-\frac{\lambda^{2}T^{4}}{16m^{4}a^{4}}\big)\big(1+\lambda \theta^{2}\pi^{4}T^{4}\big)}}
\end{align}
The above equation is written in terms of redefined variables : $y=\frac{u}{u_{0}}\;;\;a=\frac{u_{0}}{u_{B}}$, the parameter $m$ is the mass at zero temperature.
The inter-quark potential at finite temperature for noncommutative theories is obtained from \eqref{thNG} and \eqref{thscreening} to be,
\begin{align}
 \label{thpotential}
 U_{T}(L_{T}(a))=ma\int_{1}^{\frac{1}{a}} dy \sqrt{\frac{y^{4}-\frac{\lambda^{2}T^{4}}{16m^{4}a^{4}}}{y^{4}-1}}\sqrt{\frac{1+\frac{16\pi^{4}m^{4}
 \theta^{2}a^{4}}{\lambda}}{1+\lambda \theta^{2}\pi^{4}T^{4}}}
\end{align}
It is not possible to integrate the above two equations analytically. The separation however due to the presence of a finite temperature the inter-quark potential ceases to be 
coulombic even at large inter-quark separation which can be explicitly checked by computing \eqref{thpotential} for small temperature using binomial approximation. 
This phenomenon can be attributed to the breakdown of conformal symmetry (for the commutative case) in finite temperature. 

In presence of an external electric field $E$, the effective potential experienced by the $q\bar{q}$ pairs gets modified to
\begin{align}
 \label{theffective}
 &U_{T,eff}(L_{T}(a))=U_{T}(L_{T}(a))-E\cdot L_{T}=(1-R)\;\mathcal{E}_{th}\cdot L_{T}(a)+\frac{\sqrt{\lambda}}{2\pi}H(L_{T}(a))\;\;\;;\;R=\frac{E}{\mathcal{E}_{th}}
 \nonumber\\
 &H(a)=\int_{1}^{\frac{1}{a}} dy\frac{1}{\sqrt{(y^{4}-1)(1+\lambda \theta^{2}\pi^{4}T^{4})}}\Bigg[\frac{2\pi m}{\sqrt{\lambda}}a\sqrt{\big(y^{4}
 -\frac{\lambda^{2}T^{4}} {16m^{4}a^{4}}\big)\big(1+\frac{16\pi^{4}m^{4}\theta^{2}a^{4}}{\lambda}\big)}\nonumber\\
 &\;\;\;\;\;\;\;\;\;\;\;\;\;\;\;\;\;\;\;\;\;\;\;\;\;\;\;\;\;\;\;\;\;\;\;\;\;\;\;\;\;\;\;\;\;\;\;\;\;\;\;\;\;\;\;\;\;\;\;\;\;-\frac{\mathcal{E}_{th}}{am}\cdot\frac
 {\sqrt{1-\frac{\lambda^{2}T^{4}}{16m^{4}a^{4}}}\big(1+\frac{16\pi^{4}m^{4}\theta^{2}a^{4}}{\lambda}y^{4}\big)}{\sqrt{y^{4}-\frac{\lambda^{2}T^{4}}
 {16m^{4}a^{4}}}}\Bigg]
\end{align}
In the first step we have added and subtracted the term "$\mathcal{E}_{th}\cdot L_{T}$" where $\mathcal{E}_{th}$, the threshold electric field at finite temperature
 is to be found out. The slope of the potential profile \eqref{theffective} for fixed values of the physical parameters is given by,
 \begin{align}
  \label{thslope}
  \frac{dU_{T,eff}(a)}{dL_{T}}=(1-R)\;\mathcal{E}_{th}+\frac{\sqrt{\lambda}}{2\pi}\cdot \frac{dH(a)}{da}\cdot \Big(\frac{dL_{T}(a)}{da}\Big)^{-1}
 \end{align}
 At the threshold point where $R=1$, the first term vanishes and one is left with the second term alone which itself consists of the threshold value 
 \eqref{theffective}. However at the threshold point the slope the potential should be negative for all allowed values of the parameter $a$ (and henceforth the 
 separation $L_{T}$). 
 An explicit calculation leads to
 \begin{align}
  \label{thlengthslope}
  \frac{dL_{T}(a)}{da}=\frac{\sqrt{\lambda}}{2\pi m}\frac{1}{a^{2}}\frac{1}{\sqrt{1+\lambda \theta^{2}\pi^{4}T^{4}}}\Bigg[\frac{2}{\sqrt{1-\frac{\lambda^{2}T^{4}}
  {16m^{4}a^{4}}}}&\int_{1}^{\frac{1}{a}}
   dy\frac{\sqrt{y^{4}-\frac{\lambda^{2}T^{4}}{16m^{4}a^{4}}}\Big(1+\frac{16\pi^{4}m^{4}\theta^{2}a^{4}}{\lambda}y^{4}\Big)}{\sqrt{(y^{4}-1)^{3}}}\nonumber\\
   &-\bigg(\frac{1+\frac{16\pi^{4}m^{4}\theta^{2}a^{4}}{\lambda}}{\sqrt{(1+\varepsilon)^{4}-1}}\bigg)\;\Bigg]
 \end{align}
Similarly (after quite some algebraic manipulations) one finds , 
\begin{align}
 \label{thpotentialslope}
 &\;\;\;\;\;\;\frac{dH(a)}{da}=\bigg(\frac{4\pi^{2}m^{2}}{\lambda}a^{2}\sqrt{\frac{1-\frac{\lambda^{2}T^{4}}
  {16m^{4}a^{4}}}{1+\frac{16\pi^{4}m^{4}\theta^{2}}{\lambda}a^{4}}}-\frac{2\pi}{\sqrt{\lambda}}\mathcal{E}_{th}\bigg)\frac{dL_{T}(a)}{da}\\
  \label{thforce}
  &\Rightarrow \frac{dU_{T,eff}(a)}{dL_{T}}=(1-R)\;\mathcal{E}_{th}+\Bigg(\frac{2\pi m^{2}}{\sqrt{\lambda}}a^{2}\sqrt{\frac{1-\frac{\lambda^{2}T^{4}}
  {16m^{4}a^{4}}}{1+\frac{16\pi^{4}m^{4}\theta^{2}}{\lambda}a^{4}}}-\mathcal{E}_{th}\Bigg)
\end{align}
It is clear from \eqref{thscreening} that as the parameter $a\rightarrow 1$ the inter-quark separation $L_{T}\rightarrow 0$. Moreover at $a=1$ the effective potential 
\eqref{theffective} vanishes too. At $L_{T}=0$ ($a=1$) the effective force \eqref{thforce} on the $q\bar{q}$ pairs should be zero at the threshold condition. Thus,
\begin{align}
 \label{ththreshold}
 \mathcal{E}_{th}=\frac{2\pi m^{2}}{\sqrt{\lambda}}\sqrt{\frac{1-\frac{\lambda^{2}T^{4}}
  {16m^{4}}}{1+\frac{16\pi^{4}m^{4}\theta^{2}}{\lambda}}}
\end{align}
Thus we see that effect of finite temperature is to decrease the threshold electric field. However noncommutative effects don't mix up with influence of 
finite temperature (in the sense that there is no "$\theta T$" mixed terms in the expression of the threshold electric field). Similar inference can be drawn from
studying the quasi-normal modes of scalar perturbations in presence of noncommutativity as in
\cite{Edalati:2012jj}\footnote{We thank Juan F. Pedraza for bringing this to our notice.} which shows enhancement of the dissipation rate in accordance to decrement of threshold field. Also as shown in
\cite{Fischler:2012ff}\cite{Matsuo:2006ws} the NCYM is less viscous than its commutative counterpart. It is of interest to wonder whether viscosity and threshold electric field has anything to do with each other.
From \eqref{ththreshold} and \eqref{thforce} we have,
\begin{align}
 \label{thvacdecay}
 \frac{dU_{T,eff}(a)}{dL_{T}}=(1-R)\;\mathcal{E}_{th}-\frac{2\pi m^{2}}{\sqrt{\lambda}}\Bigg(\sqrt{\frac{1-\frac{\lambda^{2}T^{4}}
  {16m^{4}}}{1+\frac{16\pi^{4}m^{4}\theta^{2}}{\lambda}}}-a^{2}\sqrt{\frac{1-\frac{\lambda^{2}T^{4}}
  {16m^{4}a^{4}}}{1+\frac{16\pi^{4}m^{4}\theta^{2}}{\lambda}a^{4}}} \Bigg)
\end{align}
The monotonic nature of the second term w.r.t. the parameter a is quite clear in the allowed range of $a$. Thus the value of $\mathcal{E}_{th}$ so found suffices to cause vacuum decay
at the threshold point (R=1).
\section{Pair Production Rate of Noncommutative Schwinger Effect}
In this section we would like to estimate the rate of production of $q\bar{q}$ pairs interacting with  NCYM in presence of an external electric field along a 
noncommutative direction. As indicated in 
\eqref{rateloop} the production rate is proportional to the Wilson loop of the classical euclidean trajectory of particles under the presence of the electric 
field i.e. a circle in the $x_{0}-x_{3}$ plane. An explicit solution of the circular string profile for the gravity dual of $\mathcal{N}=4$ SYM is given in 
\cite{Berenstein:1998ij} and in \cite{Semenoff:2011ng}. Here we state the same for later purposes.
\begin{align}
 \label{crczeo}
 x_{0}(t,s)=R\;\frac{\cosh s_{0}}{\cosh s}\;\cos t \;\;\;;\;\;\;x_{3}(t,s)=R\;\frac{\cosh s_{0}}{\cosh s} \;\sin t\;\;\;;\;\;\;u(t,s)=
 u_{B}\;\frac{\tanh s_{0}}{\tanh s}
\end{align}
The solution holds true in the conformal gauge of the Polyakov action which is equivalent to the Nambu Goto action at the classical level. In 
the above $R$ indicates
the radius of the Wilson loop on the probe brane. The parameter $s$ in one of 
the co-ordinate of the 2
dimensional string worldsheet and its value on the probe brane is given by $s_{0}$, $t$ parametrizes the circular contour on the probe brane and thus has  
range $[0, 2\pi]$.
Moreover one can obtain the relation, $\sinh s_{0}=\frac{1}{Ru_{B}}$ which connects the allowed range of the worldsheet parameter to the physical
quantities like mass 
and external electric field. For the present purpose the relevant gravity dual is given by \eqref{dual}. The Polakov action in conformal gauge looks
\begin{align}
 \label{Polyakov}
 S=\frac{\sqrt{\lambda}}{4\pi}\int dtds\; \Big[U^{2}\partial_{a}X_{0}\partial_{a}X_{0}+\frac{1}{U^{2}}\partial_{a}U\partial_{a}U+
 \frac{U^{2}}{1+\alpha U^{4}}\partial_{a}X_{3}\partial_{a}X_{3}+...\Big]
\end{align}
Both the worldsheet and target spacetime have been continued to euclidean signature. We have redefined $\alpha \equiv \lambda \theta^{2}$ and have neglected the terms
involving $X_{1,2}$ . The equations of motion corresponding to \eqref{Polyakov} are given by
\begin{align}
 \label{Polyakoveqn1}
 &2\partial_{t}U\partial_{t}X_{0}+2\partial_{s}U\partial_{s}X_{0}+U(\partial_{t}^{2}X_{0}+\partial_{s}^{2}X_{0})=0\\
\label{Polyakoveqn2}
 & U(1+\alpha U^{4})(\partial_{t}^{2}X_{3}+\partial^{2}_{s}X_{3})=2(\alpha U^{4}-1)(\partial_{t}U\partial_{t}X_{3}+\partial_{s}U\partial_{s}X_{3})\\
 \label{Polyakoveqn3}
 &(\partial_{t}X_{0})^{2}+(\partial_{s}X_{0})^{2}+\frac{(1-\alpha U^{4})}{(1+\alpha U^{4})^{2}}\big((\partial_{t}X_{3})^{2}+(\partial_{s}X_{3})^{2}\big)\nonumber\\
 &+\frac{1}{U^{4}}\big((\partial_{s}U)^{2}+(\partial_{t}U)^{2}-U\partial_{t}^{2}U-U\partial_{s}^{2}U\big)=0
\end{align}
These equations are to be supplemented with the condition  $X_{0}^{2}(t,s_{0})+X_{3}^{2}(t,s_{0})=R^{2}$. The set of equations in
\eqref{Polyakoveqn1}-\eqref{Polyakoveqn3} form a system of coupled second order non-linear differential equations and in general is impossible to solve. In the
context of Gauge/String duality, solution of Wilson loops in general background has been an perplexing issue . A certain way has been suggested 
in \cite{Allahbakhshi:2013rda} based on employing a "circular ansatz", but it can be checked that such methods are valid only if the background 
has $SO(2)$ isometries in the plane of the Wilson loop (which in our case i.e. $x_{0}-x_{3}$ plane is absent). However if relevant the background is a continuous parametric deformation of AdS one can describe the
string profile as $X_{\mu}(t,s;\sigma_{i})$ where $\sigma_{i}$ collectively indicates  the deformation parameters. Expanding in power
series, $X_{\mu}(t,s;\sigma_{i})=X_{\mu}(t,s;\sigma_{i}=0)-\sigma_{i}\cdot\partial_{\sigma_{i}}X_{\mu}(t,s;\sigma_{i}=0)+\mathcal{O}(\sigma_{i}^{2})$
and noting that $X_{\mu}(t,s;\sigma_{i}=0)$ is the known AdS solution the nonlinear equations become simplified. In the present context the deformation
parameter is $\sigma \equiv \alpha=\lambda \theta^{2}$.\footnote{Crudely speaking this amounts to treating NCYM as a perturbation over YM.}
Using the expansion
\begin{align}
 \label{linear1}
 &X_{0}(t,s)=K\big(\cos t \sech s-\alpha \chi_{0}(t,s)\big)\nonumber\\&X_{3}(t,s)=K\big(\sin t \sech s-\alpha \chi_{3}(t,s)\big)\nonumber\\&U(t,s)=
 \frac{1}{K}\big(\coth s
 -\alpha \xi(t,s)\big)
\end{align}
One obtains the following equations at order $\mathcal{O}(\alpha)$
\begin{align}
 \label{Lin1}
 &\coth s\;(\partial_{t}^{2}\chi_{0}+\partial_{s}^{2}\chi_{0})-2\sin t \sech s\;\partial_{t}\xi-2\xi \cos t \sech ^{3}s\nonumber\\&-
 2 \csch ^{2}s\;\partial_{s}\chi_{0}-
 2\cos t \sech s\;\partial_{s}\xi=0\\
 \nonumber\\
 \label{Lin2}
 &\coth s\;(\partial_{t}^{2}\chi_{3}+\partial_{s}^{2}\chi_{3})-2\xi \sin t\;\sech ^{3}s+2\cos t \sech s\;\partial_{t}\xi\nonumber\\&
 -2\sin t \tanh s \sech s\;\partial_{s}\xi-
 2\csch ^{2}s\;\partial_{s}\chi_{3}=\frac{4}{K^{4}}\sin t \coth ^{2}s \csch ^{3}s\\
 \nonumber\\
 \label{Lin3}
 &\frac{3}{K^{4}}\coth ^{8}s\;(\sech ^{2}s \cos ^{2}t+\sin ^{2}t \tanh ^{2}s \sech ^{2}s)=4\coth ^{3}s \sech ^{2}s\;(1+\tanh ^{2}s)\xi\;\nonumber\\
 &-\coth s\;(\partial_{t}^{2}\xi+\partial_{s}^{2}\xi+2\frac{\csch ^{2}s}{\coth s}\partial_{s}\xi+2\csch ^{2}s\;\xi)\nonumber\\
 &-2\coth ^{4}s \sech s\;(\sin t\;\partial_{t}\chi_{0}+\cos t \tanh s\;\partial_{s}\chi_{0}+\sin t \tanh s\;\partial_{s}\chi_{3}-\cos t\;\partial_{t}\chi_{3})
\end{align}

In deriving the above, \eqref{Polyakoveqn1}-\eqref{Polyakoveqn3} has been linearized using\footnote{This is possible because in the present case we have an upper 
bound of $U=u_{B}$, in using the binomial expansion we have assumed $\lambda\theta^{2}u_{B}^{4}$to be small.}
$\frac{1}{1+\alpha U^{4}}\approx 1-\alpha U^{4}$ and then \eqref{linear1} has been used keeping in mind that terms of $\mathcal{O}(\alpha^{0})$ are AdS equations
which are
automatically zero. Moreover we have assumed $\alpha U^{4}(t,s)\approx \frac{\alpha}{K^{4}}\coth ^{4}s$ up-to first order in $\alpha$. Equations
\eqref{Lin1}-\eqref{Lin3} though being simplified than before are still daunting. Using the ansatz $\xi(t,s)=\xi(s),\;\chi_{0}(t,s)=\chi_{0}(s) \cos t,\;\chi_{3}(t,s)=
\chi_{3}(s) \sin t$ in \eqref{Lin1}and
\eqref{Lin2} one has
\begin{align}
\label{ALin1}
&\partial_{s}\chi_{0}=\frac{1}{2}\sinh ^{2}s \coth s\;(\partial_{s}^{2}\chi_{0}-\chi_{0})-\xi \sinh ^{2}s \sech ^{3}s -
\sech s \tanh s \sinh ^{2}s\;\partial_{s}\xi\\
\label{ALin2}
&\partial_{s}\chi_{3}=\frac{1}{2}\sinh ^{2}s \coth s\;(\partial_{s}^{2}\chi_{3}-\chi_{3})-\xi \sinh ^{2}s \sech ^{3}s
-\sech s \tanh s \sinh ^{2}s\;\partial_{s}\xi\nonumber\\
&\;\;\;\;\;\;\;\;\;\;-\frac{2}{K^{4}}\coth ^{2}s \csch s
\end{align}
Since the set \eqref{Lin1}-\eqref{Lin3} are coupled differential equations the solution of the first two has to satisfy the other one. Substituting \eqref{ALin1},
\eqref{ALin2} in \eqref{Lin3} and noting that the resulting equation has to be satisfied for all values of parameter $t$ one obtains the following three equations
\begin{align}
\label{Utility1}
&\tanh s \sinh s\;\partial_{s}^{2}\chi_{0}-2\sech s\;\chi_{3}-\tanh s \sinh s\;\chi_{0}+\frac{3}{K^{4}}\coth ^{2}s \csch^{2}s=0\\
\label{Utility2}
&\tanh s \sinh s\;\partial_{s}^{2}\chi_{3}-2\sech s\;\chi_{0}-\tanh s \sinh s\;\chi_{3}-\frac{1}{K^{4}}\csch ^{2}s=0\\
\label{Radial}
&\coth s\;\partial_{s}^{2}\xi-2\xi \coth s \csch ^{2}s\;(1+3\tanh ^{2}s)+2\;(\csch ^{2}s-1)\;\partial_{s}\xi=0
\end{align}
It can be checked that \eqref{Radial} has no real solution, further more \eqref{Radial} being a linear equation permits a 
solution of the form $\xi=0$. Thus we are left 
with the first two equations \eqref{Utility1},\eqref{Utility2} which are coupled differential equations themselves. To simplify 
those we define the following variables
whose significance is rather obscure.
\begin{align}
 \label{variables}
 &\chi_{+}(s)=\chi_{0}(s)+\chi_{3}(s)\;\;\;\;\;\;;\;\;\;\;\;\;\chi_{-}(s)=\chi_{0}(s)-\chi_{3}(s)
\end{align}
In terms of the above one has
\begin{align}
 \label{key1}
 &\partial_{s}^{2}\chi_{+}-\chi_{+}-2\csch ^{2}s \;\chi_{+}+\frac{\csch ^{3}s}{K^{4}}\coth s\;(3\coth ^{2}s-1)\\
 \label{key2}
  &\partial_{s}^{2}\chi_{-}-\chi_{-}+2\csch ^{2}s \;\chi_{-}+\frac{\csch ^{3}s}{K^{4}}\coth s\;(3\coth ^{2}s+1)
 \end{align}
Digressing  a bit from the main discussion let us see the first order correction to the on-shell action in light of the perturbation theory set up. From the 
decomposition \eqref{linear1} one has up-to $\mathcal{O}(\alpha)$
\begin{align}
&(\partial_{t}X_{0})^{2}=K^{2}\sin ^{2}t\;(\sech ^{2}s-2\alpha \sech s\;\chi_{0})\\
&(\partial_{t}X_{3})^{2}=K^{2}\cos ^{2}t\;(\sech ^{2}s-2\alpha \sech s\;\chi_{3})\\
&(\partial_{s}X_{0})^{2}=K^{2}\cos ^{2}t\;(\sech ^{2}s \tanh ^{2}s+2\alpha \sech s \tanh s\;\partial_{s}\chi_{0})\\
&(\partial_{s}X_{3})^{2}=K^{2}\sin ^{2}t\;(\sech ^{2}s \tanh ^{2}s+2\alpha \sech s \tanh s\;\partial_{s}\chi_{3})
\end{align}
Using  the above in the Polyakov action in presence of the NC dual \eqref{Polyakov} and approximating 
$\frac{1}{1+\frac{\alpha}{K^{4}}\coth ^{4}s}\approx 1-\frac{\alpha}{K^{4}}\coth ^{4}s$ one has in the first order of the effective 
parameter.
\begin{align}
 \label{Action1}
 &\mathcal{L}_{on shell}=2\csch ^{2}s-2\alpha \frac{\cosh s}{\sinh ^{2}s}\bigg[\sin ^{2}t \Big\{\chi_{0} -\tanh s\;\partial_{s}\chi_{3}-
 \frac{\coth ^{2}s}{2K^{4}} \sech s \Big\}\nonumber\\
&\;\;\;\;\;\;\;\;\;\;\;\;\;\;\;\;\;\;\;\;\;\;\;\;\;\;\;\;\;\;\;\;\;\;\;\;\;\;\;+\cos ^{2}t \Big\{\chi_{3}-\tanh s\;\partial_{s}\chi_{0}
-\frac{\coth ^{4}s}{2K^{4}} \sech s\Big\}\bigg]
\end{align}
Using the equations \eqref{Utility1},\eqref{Utility2} and the fact that $\xi(s)=0$ one can reduce the first order correction to above expression to 
\begin{align}
\label{Action2}
&\delta_{\alpha}\mathcal{L}_{on shell}=-\alpha \cosh s\bigg[\sin ^{2}t\Big\{2\csch^{2}s\;\chi_{0}+\frac{3}{K^{4}}\coth ^{2}s \sech s \csch ^{2}s-
\partial_{s}^{2}\chi_{3}+\chi_{3}\Big\}\nonumber\\
&\;\;\;\;\;\;\;\;\;\;\;\;\;\;\;\;\;\;\;\;\;\;\;\;\;\;\;\;\;\;\;\;+\cos ^{2}t \Big\{2\csch ^{2}s\;\chi_{3}-\frac{1}{K^{4}}\coth^{4}s \sech s \csch ^{2}s
-\partial_{s}^{2}\chi_{0}+\chi_{0}\Big\}\bigg]\nonumber\\
&\;\;\;\;\;\;\;\;\;\;\;\;\;\;\;=-\frac{2\alpha}{K^{4}}\coth ^{2} \csch ^{2}s \big[1+ \csch ^{2} s \cos ^{2} t\big]
\end{align}
In the last line \eqref{Utility1},\eqref{Utility2} have been put to use. Thus we see that the equations of motions alone determine the first order 
correction of the on-shell action from the commutative counterpart.
In the context of holographic entanglement entropy similar methods have been presented in \cite{Allahbakhshi:2013rda}. For finding out the limit of the 
integration and its connection to physical variables one has to solve the equations of motion.
Returning to our main discussion, the real part of the solution of \eqref{key1} is 
\begin{align}
 \label{sol}
 \chi_{+}(s)=\chi_{0}(s)+\chi_{3}(s)=-\frac{1}{6K^{4}}\coth ^{4}s \sech s
\end{align}
The equation \eqref{key2} which dictates the deviation of the circular symmetry cannot be solved by analytical means. It is worthwhile to note 
that \eqref{key2} is not a linear equation and  does not admit
a solution $\chi_{-}(s)=0$. However since it is a second order differential equation with two constants one can redefine them to set $\chi_{-}(s=s_{0})=0$ for a
specific value $s_{0}$, which is essentially imposing the boundary condition of the loop being circular at the probe brane .\footnote{There exists no known methods to solve
a generic second order partial differential equation. We don't claim that
the first derivative of $\chi_{-}$ is zero at $s_{0}$} 
Thus at the point given by $s=s_{0}$ the profile is circular and the variable $\chi_{+}$ is twice the radius of the loop($R$). Putting the above 
in mathematical language, from \eqref{linear1} and \eqref{sol}
\begin{align}
 \label{boundary}
 &R=K(\sech s_{0}+\frac{\alpha}{12K^{4}}\coth ^{4}s_{0} \sech s_{0})\nonumber\\
 &u_{B}=\frac{1}{K}\coth s_{0}
\end{align}
From the above one gets, 
\begin{align}
 Ru_{B}=\csch s_{0} \big(1+\frac{\alpha}{12}u_{B}^{4}\big)
\end{align}
This relation serves to define the integration limits and also
connects the on-shell value of the action to physical parameters. From the above one has in first order of the noncommutative parameter,
\begin{align}
 \coth s_{0}=\sqrt{1+R^{2}u_{B}^{2}(1-\frac{\alpha}{6}u_{B}^{4})} 
\end{align}
In presence of an external electric field in the $x_{3}$ direction the on shell value of
the action gets modified to\footnote{U(1) gauge fields contribute to the
string Lagrangian via a boundary term. For constant electromagnetic field the string equations of motion are unchanged but the boundary
conditions are altered (Robin).
For inhomogeneous fields this is not the case. Schwinger Effect for inhomogeneous fields have been explored via holographic methods 
in \cite{Lan:2018xnq}}
\begin{align}
 \label{onshell}
S_{onshell}=\sqrt{\lambda}\bigg[\Big\{(1-3\eta)\coth s_{0}-1\Big\}-5\eta\frac{1}{\coth s_{0}}+8\eta\frac{1}{\coth ^{4}s_{0}}
 -\mathcal{E}\csch ^{2}s_{0}\bigg]
\end{align}
In the above we replaced the constants by hyperbolic functions \eqref{boundary} and have defined 
$\eta=\frac{\alpha}{30}u_{B}^{4} \equiv \frac{\lambda \theta^{2}}{30}u_{B}^{4}$ and $\mathcal{E}=\frac{\pi}{\sqrt{\lambda}u_{B}^{2}}(1+\frac{5}{2}\eta)^{2}E$. Note
that the dependence on the radius ($R$) is now encoded in the hyperbolic functions themselves. Quite similar to arguments in \cite{Semenoff:2011ng},
\cite{Bolognesi:2012gr} at large value of R ( large E) the production rate \eqref{rateloop},\eqref{wloopstring},\eqref{onshell} is dominated
by $\Gamma \sim  \exp (\sqrt{\lambda} R^{2}E) $ similar to the phase (in potential analysis) when $V_{sch}$  does not permit a tunneling barrier. However
for small  R ($\pi R^{2} E$ not dominating the other terms) the  approximate production rate is 
\begin{align}
 \label{estimation}
 \Gamma \sim \exp (-S_{onshell}) \sim \exp\big(-\frac{\sqrt{\lambda}u_{B}^{2}}{2}(1-5\eta)(1-30\eta)R^{2}+\pi ER^{2}+\mathcal{O}(R^{4}) \big)
\end{align}
It is evident that as R varies one moves from a tunneling or damped production phase (when the first term in \eqref{estimation} dominates) to a spontaneous 
production 
phase (when the second term dominates). This is quite synonymous to the potential analysis with the identification $\Gamma \sim \exp (-V_{effective})$.
The potential barrier vanishes when both terms in \eqref{estimation} cancels each other which happens at 
\begin{align}
 \label{estimation2}
 E_{threshold}\sim \frac{\sqrt{\lambda}}{2\pi}u_{B}^{2}(1-5\eta)(1-30\eta) \sim \frac{2\pi m^{2}}{\sqrt{\lambda}}\Big(1-\frac{56}{3}\frac{\pi^{4}m^{4}\theta^{2}}
{\lambda}\Big)
\end{align}
To get the production rate one needs to extremise the on-shell action with respect to R for reasons mentioned before. Instead of 
extremizing w.r.t. R one can
extremise w.r.t. $\csch s_{0}$, however doing so one is left to solve a sextic equation. To simplify the situation note that the value
of $\coth s_{0}$ is proportional
to the mass of the quark (W boson) via the relation derived earlier . Thus for heavy mass, (actually $\frac{\lambda \theta^{2}m^{4}}{m^{6}}\ll1$) the
contribution
of the second and third term of \eqref{onshell} is negligible. Under those circumstances one has
\begin{align}
 \label{redefonshell}
 S_{onshell}\approx \sqrt{\lambda}\big[(1-3\eta)\coth s_{0}-1-\mathcal{E}\csch ^{2}s_{0}\big]
\end{align}
Extremizing the above w.r.t. $\csch s_{0}$ (i.e. R) one is lead to $\coth s_{0}=\frac{1-3\eta}{2\mathcal{E}}$ which is a condition connecting R 
and E. From the relation
thus obtained the on shell value of the action is
\begin{align}
 \label{redefonshellvalue}
 S_{onshell}=\frac{\sqrt{\lambda}\big(1-3\eta\big)}{2}\bigg[\frac{\big(1-3\eta\big)\big(1-5\eta\big)\sqrt{\lambda}u_{B}^{2}}{2\pi E}-\frac{2}{\big(1-3\eta\big)}
 +\frac{2\pi E}{\big(1-3\eta\big)\big(1-5\eta\big)\sqrt{\lambda}u_{B}^{2}}\bigg]
\end{align}
Thus the production rate which is proportional to the (negative) exponential of the on-shell action \eqref{rateloop},\eqref{wloopstring},\eqref{redefonshellvalue}
is given by 
\begin{align}
 \label{final}
\Gamma \sim \exp\bigg[-\frac{\sqrt{\lambda}}{2}\Big(1-\frac{8\pi^{4}m^{4}\theta^{2}}{5\lambda}\Big) \Big\{\sqrt{\frac{\mathcal{E}_{T}}{E}}-
  \sqrt{\frac{E}{\mathcal{E}_{T}}}  \Big\}^{2} + \frac{8\pi^{4}m^{4}\theta^{2}}{5\sqrt{\lambda}} \bigg]
\end{align}
Where we have restored the physical parameters via the relation $\eta=\frac{8\pi^{4}m^{4}\theta^{2}}{15 \lambda}$. In the above the threshold electric
field is given 
by 
\begin{align}
 \label{threshold}
 \mathcal{E}_{T}=\frac{2\pi m^{2}}{\sqrt{\lambda}}\bigg(1-\frac{8}{5}\frac{\pi^{4}m^{4}\theta^{2}}{\lambda}\bigg)\bigg(1-\frac{8}{3}\frac{\pi^{4}m^{4}\theta^{2}}
 {\lambda}\bigg) \approx \frac{2\pi m^{2}}{\sqrt{\lambda}}\bigg(1-\frac{64}{15}\frac{\pi^{4}m^{4}\theta^{2}}{\lambda}\bigg)
\end{align}
At low electric field , $E\ll \mathcal{E}_{T}$, the second term in \eqref{final} ceases to contribute and one is left with
\begin{align}
 \label{lowenergy}
 \Gamma \sim \exp \bigg[-\frac{\pi m^{2}}{E}\bigg(1-\frac{88\pi^{4}m^{4}\theta^{2}}{15\lambda}\bigg)+
 \Big(\frac{8\pi^{4}m^{4}\theta^{2}}{5\sqrt{\lambda}}\Big)\bigg]
\end{align}
One can compare \eqref{lowenergy} to the result of \cite{Chair:2000vb}, and both of them show the same pattern with the identification
of $\overrightarrow{B} \sim \theta$ where $\overrightarrow{B}$ indicates an external magnetic field in \cite{Chair:2000vb}. In presence 
of strong magnetic fields commutativity is lost and the theory is described by non-commutative physics\cite{Miransky:2015ava}. We come up with "three" values of threshold electric field in 
\eqref{critical}\eqref{estimation2}\eqref{threshold}. The reason they don't match is because the later two values have been found via perturbative methods while the former \eqref{critical} by exact methods. Moreover
one should notice that\eqref{estimation2} is estimated from \eqref{estimation} which itself is valid for small R, unlike \eqref{threshold} which is derived at the limit of large quark mass (where the assumption on R is relaxed).
However all of them point towards the same fact, \textit{threshold electric field is decreased due to noncommutativity}.
 A reason for concern may be the extra ( $\theta$ dependent) term 
in \eqref{final} and \eqref{lowenergy} which is independent of the electric field. We believe this is an effect of our simplification of solving a quadratic equation 
instead of a sextic one (see above).

\textit{A reason for concern:} One can also find out the "threshold" electric field from the DBI action ( by claiming that the DBI action should be real for all allowed values of the electric field ) \cite{Wu:2015krf}
\cite{Hashimoto:2013mua}. However such analysis shows that the threshold electric field is the same as that of the commutative case irrespective of whether the applied electric field is parallel or perpendicular to the
non-commutative directions contrary to our findings. This issue is not clear to us at the moment.

\section{Conclusions}

In this paper we have performed an inter-quark potential analysis to find the effective potential barrier in presence of an external electric field in noncommutative
gauge theory. From the same we have shown that the threshold electric field is decreased from it's commutative counterpart. In presence of noncommutativity there exist
strong repulsive forces between the particles at short distances i.e. the coulombic interaction develops a short distance repulsive correction. This implies
the electrostatic potential energy needed to tear out the virtual particles is less than usual explaining the result found. We have also argued that noncommutativity
does not lead to confinement as at large distances the behavior of the potential is essentially coulombic as demonstrated. We also have found out the thermal 
corrections to the above and have seen that finite temperature effects don't entangle with (space-space) noncommutative ones as expected which also can be seen by putting
the values of \eqref{dual}\eqref{thdual}in 
\cite{Sato:2013hyw}. Moreover, we have also perturbatively 
computed the corrections to circular Wilson loop over the known commutative result in the first order of the noncommutative deformation parameter, and hence the 
decay rate has been found out from which the decrement 
of the threshold value is also clear. To the best of our knowledge such an analysis for Noncommutative Holographic Schwinger effect is new . At low electric field our
result shares the same pattern with that of \cite{Chair:2000vb} for noncommutative 
U(1) gauge theory.

\acknowledgments

UNC would like to thank Shibaji Roy for discussions during this work.


\bibliographystyle{JHEP}
\bibliography{jhep}
\end{document}